\documentclass[preprint2]{aastex}

\usepackage{epsf}
\usepackage{graphics}
\usepackage{color} 
\usepackage{lscape}
\usepackage{longtable}

\newcommand{\be}{\begin{equation}}
\newcommand{\ee}{\end{equation}}

\newcommand{\bx}{$\beta_{\rm X}$}
\newcommand{\aox}{$\beta_{\rm ox}$}

\newcommand{\rb}[1]{\raisebox{1.5ex}[-1.5ex]{#1}}

\newcommand{\plm}{$\pm$}
\newcommand{\nh}{$N_{\rm H}$}
\newcommand{\swift}{\mbox{\it Swift}}	    % defines Chandra name
\newcommand{\chandra}{\mbox{\it Chandra}}	    % defines Chandra name
\newcommand{\xmm}{\mbox{\it XMM-Newton}}	    % defines Chandra name
\newcommand{\meszaros}{M\'esz\'aros~}
\newcommand{\days}{125 }
\shorttitle{Swift and XMM Observations of GRB\,060729}
\shortauthors{Grupe et al.}

\begin{document}

%\input DGrupe_clipfig.tex
%\useunitmm

\def\etal{{\it et\thinspace al.}\ }
\def\alp{{$\alpha$}\ }
\def\al2{{$\alpha^2$}\ }

%\def \charthoffset {\hspace{0.2cm}} \def \charthsep {\hspace{0.3cm}}
%\def \chartvsepcap {\vspace{0.3cm}}
%\def \chartvsep {\vspace{0.1cm}}
%\newcommand{\putchartb}[1]{\clipfig{#1}{75}{20}{7}{275}{192}}
%\newcommand{\putchartc}[1]{\clipfig{#1}{55}{33}{19}{275}{195}}
%
%
%
%\newcommand{\chartlineb}[2]{\parbox[t]{18cm}{\noindent\charthoffset\putchartb{#1}\charthsep\putchartb{#2}\chartvsep}}
%\newcommand{\chartlinec}[3]{\parbox[t]{18cm}{\noindent\charthoffset\putchartc{#1}\charthsep\putchartc{#2}\chartvsep\putchartc{#3}\chartvsep}}

%% LaTeX will automatically break titles if they run longer than
%% one line. However, you may use \\ to force a line break if
%% you desire.

\title{\swift\ and \xmm\ Observations of the Extraordinary GRB\,060729:
 More than 125 days of X-ray afterglow}

%% Use \author, \affil, and the \and command to format
%% author and affiliation information.
%% Note that \email has replaced the old \authoremail command
%% from AASTeX v4.0. You can use \email to mark an email address
%% anywhere in the paper, not just in the front matter.
%% As in the title, you can use \\ to force line breaks.

\author{Dirk Grupe\altaffilmark{1},\email{grupe@astro.psu.edu}
Caryl Gronwall\altaffilmark{1},
Xiang-Yu Wang\altaffilmark{1,2},
Peter W.A. Roming\altaffilmark{1},
Jay Cummings\altaffilmark{3},
Bing Zhang\altaffilmark{4},
Peter \meszaros\altaffilmark{1,5},
Maria Diaz Trigo\altaffilmark{6},
Paul T. O'Brien\altaffilmark{7},
Kim L. Page\altaffilmark{7},
Andy Beardmore\altaffilmark{7},
Olivier Godet\altaffilmark{7},
Daniel E. vanden Berk\altaffilmark{1},
Peter J. Brown\altaffilmark{1},
Scott Koch\altaffilmark{1},
David Morris\altaffilmark{1},
Michael Stroh\altaffilmark{1},
David N. Burrows\altaffilmark{1},
John A. Nousek\altaffilmark{1}, 
Margaret McMath Chester\altaffilmark{1},
Stefan Immler\altaffilmark{8,9},
Vanessa Mangano\altaffilmark{10},
Patrizia Romano\altaffilmark{11},
Guido Chincarini\altaffilmark{11},
Julian Osborne\altaffilmark{7},
Takanori Sakamoto\altaffilmark{3},
Neil Gehrels\altaffilmark{3}
}

\altaffiltext{1}{Department of Astronomy and Astrophysics, Pennsylvania State
University, 525 Davey Lab, University Park, PA 16802} 
\altaffiltext{2}{Department of Astronomy, Nanjing University, 
Nanjing 210093, China}
\altaffiltext{3}{Astrophysics Science Division, Astroparticle Physics Laboratory,
Code 661, NASA Goddard Space Flight Center, Greenbelt, MD 20771 }
\altaffiltext{4}{Department of Physics, University of Nevada, Las Vegas, 
NV 89154}
\altaffiltext{5}{Department of Physics, Pennsylvania State University,
University Park, PA 16802}
\altaffiltext{6}{XMM-Newton Science Operations Centre, ESA, 
Villafranca del Castillo, P.O. Box 50727, 28080 Madrid, Spain}
\altaffiltext{7}{Department of Physics \& Astronomy, University of Leicester,
Leicester, LE1 7RH, UK}
\altaffiltext{8}{Astrophysics Science Division, X-Ray Astrophysical Laboratory,
Code 662, NASA Goddard Space Flight Center, Greenbelt, MD 20771}
\altaffiltext{9}{Universities Space Research Association,
10211 Wincopin Circle, Columbia, MD 21044}
\altaffiltext{10}{INAF - Istituto di Astrofisica Spaziale e Fisica Cosmica di 
Palermo,  Via Ugo La Malfa 153, 90146 Palermo, Italy }
\altaffiltext{11}{INAF-Osservatorio Astronomico di Brera, via E. Bianchi 46,
I-23807 Merate, Italy}

%% Notice that each of these authors has alternate affiliations, which
%% are identified by the \altaffilmark after each name.  Specify alternate
%% affiliation information with \altaffiltext, with one command per each
%% affiliation.

%\altaffiltext{1}{Visiting Astronomer, Cerro Tololo Inter-American Observat}

%% Mark off your abstract in the ``abstract'' environment. In the manuscript
%% style, abstract will output a Received/Accepted line after the
%% title and affiliation information. No date will appear since the author
%% does not have this information. The dates will be filled in by the
%% editorial office after submission.

\begin{abstract}
We report the results of the \swift\ and \xmm\ observations of the
\swift-discovered long Gamma-Ray Burst GRB\,060729 ($T_{90}$=115s).
The afterglow of this burst was exceptionally bright in X-rays as well as at
UV/Optical wavelengths showing an unusually long slow decay phase ($\alpha$=0.14\plm0.02)
suggesting a larger energy injection phase at early times than in other bursts.
The X-ray light curve displays a break  at about 60 ks
after the burst.  The X-ray decay slope after the break is 
$\alpha$=1.29\plm0.03. Up to \days days after
the burst 
we do not detect a jet break, suggesting that the jet opening angle is larger
than  28$^{\circ}$. In the first 2 minutes after the burst (rest frame) the X-ray spectrum
of the burst changed dramatically from a hard X-ray spectrum to a very soft one. We find
that the X-ray spectra at this early phase can all be fitted by an absorbed single power
law model or alternatively by a blackbody plus power law model. The power law fits show
that the X-ray spectrum becomes steeper while the absorption column density decreases.
In the blackbody model we find that  
the temperature changes from $kT=0.6$ keV at 85 s after the burst
to 0.1 keV at 160 s after the burst in the rest frame. 
In \swift's UV/Optical telescope the afterglow
was clearly
 detected up to 9 days after the burst 
 in all 6 filters and even longer in some of the
UV filters with the latest detection in the UVW1 
31 days after the burst, 
which is one of the latest detections of an afterglow by the
UVOT. A break at about 50 ks is clearly detected in all 6 UVOT filters from 
a shallow decay slope of about 0.3 and a steeper decay
slope of 1.3. The spectral
energy distributions show that there is no spectral development between X-rays and the
UV/optical throughout the observation.
In addition to the \swift\ observations we also present and discuss the results from a
61 ks
Target-of-Opportunity observation by \xmm\ starting about 12 hours after the burst. 
These observations show a typical afterglow X-ray spectrum with \bx=1.1 and 
absorption column density of $1\times 10^{21}$ cm$^{-2}$.
\end{abstract}


\section{Introduction}

Gamma-Ray Bursts (GRBs) are the most powerful explosions in the 
present-day Universe. With
the launch of the \swift\ Gamma-Ray Burst Explorer Mission
\citep{gehrels04} 
in November 2004 a new era in GRB science has started. \swift\ is able to
observe the afterglow of a 
burst with its narrow-field instruments, the X-ray Telescope 
\citep[XRT, ][]{burrows05} and the UV/Optical telescope  
\citep[UVOT; ][]{roming05}, typically within 2 minutes after the 
detection by the Burst
Alert Telescope  \citep[BAT, ][]{barthelmy05a}. Several phenomena were
discovered by \swift, like the occurrence of giant flares during the first 1000
s after the burst \citep[e.g.][]{burrows05b, falcone06} or the canonical light curves of
GRB afterglows \citep[][]{nousek06, zhang06}. 

GRB\,060729 was discovered by \swift\ on 2006 July 29 \citep{grupe06b}
as one of the brightest bursts ever detected in X-rays
by \swift\ \citep[With the brightest GRB in XRT so far being GRB 061121;
][]{page07}.
Besides GRB\,050525A \citep[][]{blustin05},
GRB 060218 \citep{campana06b}, GRB 060614 \citep{mangano07}, and GRB 061121
\citep{page07},
GRB\,060729 is the burst with the best UVOT follow up even up to 9 days after the
burst in all 6 UVOT filters and even longer in some of the UV filters.
 In the XRT the afterglow has been detected by more than \days days
after the
burst. This is the longest follow-up observation with a detection 
of an afterglow ever performed by  \swift.
Similar coverage, but shorter, has only been performed for GRBs 050416A,
060319, and 060614. 
Even though the burst was bright in the BAT, it was too faint to be
 detected by Konus-Wind (Frederiks, priv. comm.).

Even though the sun angle was small in RA (2.2 h), 
due to the declination of Dec=--62$^{\circ}$
the afterglow was circumpolar for most southern observatories, and was observed
by the ESO VLT using FORS2 and by Gemini South using GMOS \citep{thoene06}. 
A redshift of z=0.54 was determined from the optical
spectra by \citet{thoene06}. 
GRB\,060729 was also observed
 by ROTSE IIIa, located at the Siding Spring Observatory, Australia, by
\citet{quimby06a}, who reported an initial upper limit of  16.6 mag 64 s 
after the BAT trigger. They were
able to detect the afterglow with ROTSE IIIa up to 175 ks after the 
burst when it was still
 at 19.4 mag \citep{quimby06b}, decaying with a slope $\alpha$=0.23. 
 \citet{cobb06} 
measured a decay slope in the I band
of $\alpha_{\rm I}$=1.5 based on CTIO 1.3m SMARTS observations between 4.6 to
17.6 days after the burst.  

The paper is organized as follows: In \S\ref{observe} we describe the
observations and the data reduction.  In \S\ref{results} we
present the data analysis.  The discussion of our results is given in
\S\ref{discuss}. 
Throughout
the paper decay and energy spectral indices $\alpha$ and $\beta$ 
are defined by $F_{\nu}(t,\nu)\propto
(t-t_0)^{-\alpha}\nu^{-\beta}$, with $t_0$ the trigger time of the burst. 
Luminosities are calculated assuming a $\Lambda$CDM
cosmology with $\Omega_{\rm M}$=0.27, $\Omega_{\Lambda}$=0.73 and a Hubble
constant of $H_0$=71 km s$^{-1}$ Mpc$^{-1}$ using the luminosity distances
$D_{\rm L}$
given by \citet{hogg99} resulting in $D_{\rm L}$=3120 Mpc. 
All errors are 1$\sigma$ unless stated otherwise. 

\section{\label{observe} Observations and data reduction}

The 
\swift~BAT triggered on the pre-cursor of
GRB\,060729 at 19:12:29 UT on 2006 July 29  \citep{grupe06b, grupe06a}. 
 \swift's X-ray Telescope began observing the afterglow 
124\,s after the trigger. The UVOT started the observations 135 s after the 
BAT trigger.

The \swift~XRT observed GRB\,060729 in the Windowed Timing (WT) and Photon
Counting (PC) observing modes \citep{hill04}. 
The XRT data were reduced by the {\it xrtpipeline} task version 0.10.4. 
The WT mode data at the beginning of the XRT observation had to be treated with
special care. During the first 20s after the start (130--150s after the BAT
trigger)
of the WT XRT observation the
satellite was still settling causing the target to move on the XRT CCD towards
the dead columns at DETX=319--321 \citep{abbey06}. 
In order to correct for the photon losses due
to these dead columns we measured the offset of the source at each second and
calculated a correction factor according to the losses of the WT mode Point
Spread Function (PSF).  After 150s after the trigger all WT mode data were
corrected by the same factor. Source and Background photons were selected 
by {\it XSELECT} version 2.4 in boxes with a length of 40 pixel.
For count rates $>$ 150 counts s$^{-1}$, however,
in the WT mode the data had to be corrected for pileup. 
In order to correct for
pileup in the light curve and
for the spectral analysis and the determination of the hardness 
ratio\footnote{The hardness ratio is defined by
$HR=(hard-soft)/(hard+soft)$ where {\it soft} is the counts in the 0.3-1.0 keV band and
{\it hard} is the counts in the 1.0-10.0 keV band, respectively.} we excluded the central
regions of the PSF at the source position depending on the count rate as 
described in \citet{romano06}.
For the PC mode data the source photons were selected
in a circular region with a radius
of r=59$^{''}$ and the background photons in a circular region close by with a
radius r=176$^{''}$ in the first segments. For the later data the radii were reduced 
to 47$^{''}$ and 24$^{''}$ for the source, and 137$^{''}$ and 96$^{''}$ for the
background.
For the spectral data only events with grades 0--2 and 0--12 were
selected with {\it XSELECT} for the WT and PC mode data, respectively.
Note that the source photons for the spectral analysis 
of the PC mode data of the first orbit were selected
in a ring with an inner radius of $16 \farcs 5$ and an outer radius of
 $71 \farcs 0$ in order to avoid the effects of pile-up 
 \citep[e.g. ][]{pagani06, vaughan06}.
The spectral data were re-binned by using 
{\it grppha} version 3.0.0 having 20 photons per bin. The spectra were 
analyzed with {\it XSPEC} version 12.3.0 \citep{arnaud96}. 
The auxiliary response files were created by {\it
xrtmkarf} and corrected using the exposure maps,
and the standard response matrices {\it swxwt0to2\_20010101v008.rmf}
and {\it swxpc0to12\_20010101v008.rmf}
 were used for the WT and PC mode data, respectively. All spectral fits were performed in
 the observed 0.3--10.0 keV energy band.
{\bf For the errors of the spectral fit parameters
we used the standard $\Delta
\chi^2$=2.7 in XSPEC, that is equivalent to a 90\% confidence region for a
single parameter.} 
 
Background-subtracted X-ray flux
light curves in the 0.3--10.0 keV energy range of the 
\swift\ observations were constructed using the ESO Munich
Image Data Analysis Software {\it MIDAS} (version 04Sep) and by an 
IDL program that corrects for PSF losses, in particular when the source is 
located on one of the dead columns on the XRT CCD detector. 
The light curve was binned as follows: the WT mode data with 1000 photons per bin,
and the pc mode data with  200 counts per bin in the first days after the burst and 20 or
10 at the end of the observations.
 The count rates were converted into unabsorbed
 flux units using energy conversion factors
(ECF) which were determined by calculating  the count rates and the
unabsorbed fluxes in the 0.3--10.0 keV energy band using XSPEC
as described in \citet{nousek06}. The XRT data at the beginning of the
observation show dramatic spectral changes and require specific ECFs for each
time bin. The later data, however, do agree with a typical afterglow spectrum
and the count rates were converted by one ECF=5$\times 10^{-11}$ ergs s$^{-1}$
cm$^{-2}$ (counts s$^{-1}$)$^{-1}$.

\begin{figure*}
%\epsscale{1.8}
\epsscale{1.5}
\plotone{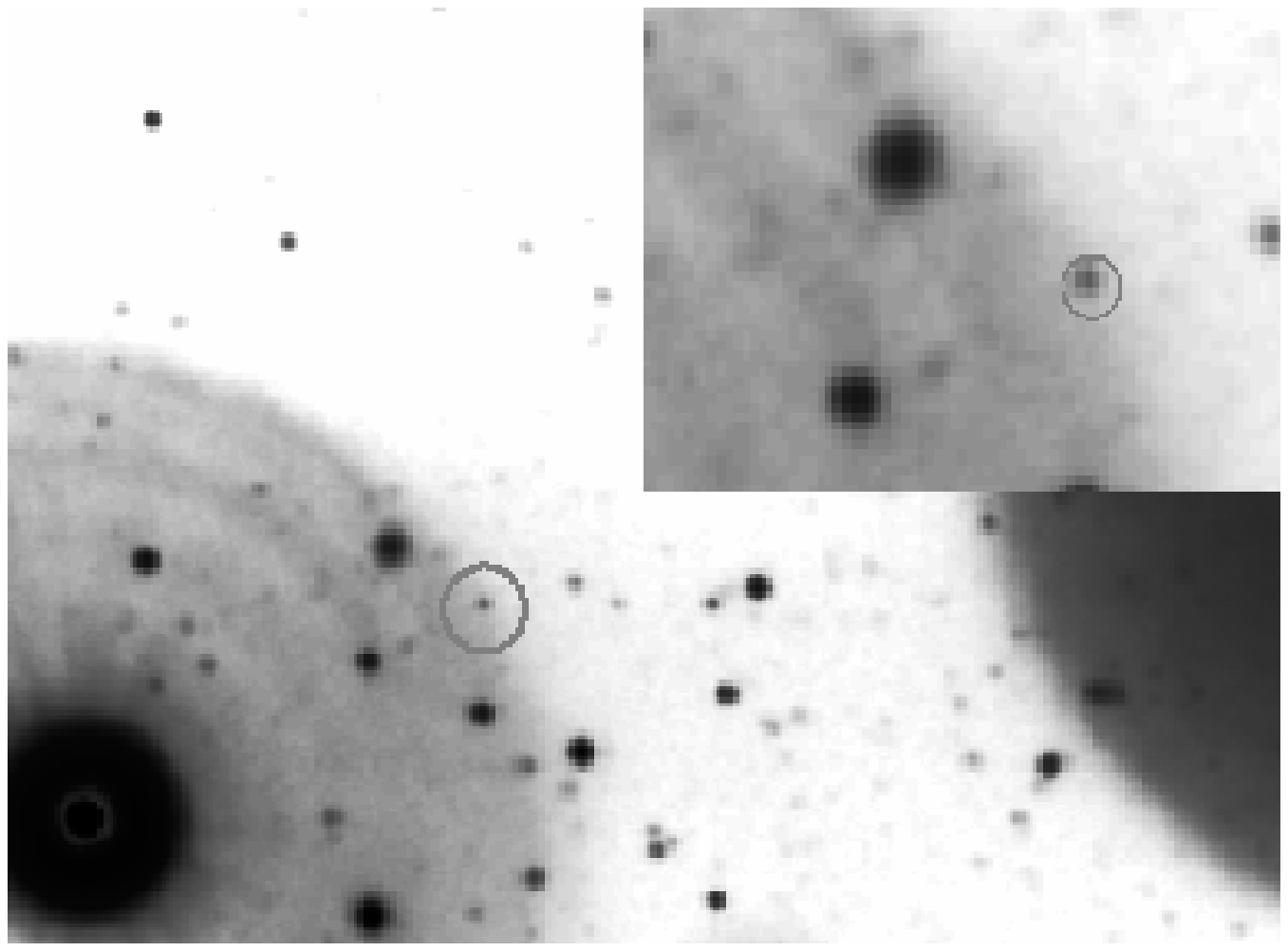}
\caption{\label{grb060729_w1_image} UVW1 image ($5 \farcm 7 \times 4 \farcm 2$)
of the field of GRB\,060729 with an exposure
of 550 ks.
The circles at the source position
displays the  XRT position as given in Section\,\ref{position}. 
The circles in the large
image show a 8$^{''}$ radius at the source
just for display purposes, and the 8$^{''}$ background extraction region.
The zoom-in 
image (80$^{''}\times 60^{''}$)
in the upper right corner shows the $3 \farcs 5$ XRT error radius. 
Note the bright star
which is only 107$^{''}$ away from the position of the afterglow of GRB\,060729.
%{\bf Due to the size restrictions on astro-ph we could not include this figure.}
}
\end{figure*}

The \swift~UVOT observations of GRB060729 began with the automated ``GRB" 
sequence, which provided finding chart images in White (100 s) and V (400 s), 
then began cycling through all 6 UV and optical filters starting 739 s after the 
trigger. The source data of these early White and V images were extracted from a 
circle with a radius of 6$^{''}$. 
GRB\,060729 remained detectable in all 6 filters for more than 9 days after the
trigger,  then for 12 days after the trigger 
in UVW2 ($\lambda_c=1930$\AA) and for 31 days after the trigger in 
UVW1 ($\lambda_c=2510$\AA). 
During the first days after the burst each observation
in each single orbit was analyzed. For the later data the images were coadded 
with {\it uvotimsum} in
order to improve the signal to noise ratio. The data were analyzed with the UVOT
software tool {\it uvotsource}. Due to the bright F3V star HD 45187
(9.4 mag in B, 
107\farcs5 away from GRB\,060729) 
extra caution had to be taken for the source extraction and the
background subtraction.
We chose a selection radius of
4$^{''}$ for the source and $8^{''}$ for the background in all filters placed at
a position nearby on the rim of the bright star's halo
as displayed in Figure\,\ref{grb060729_w1_image}.
In order to
correct for losses due to this small source extraction radius, we did an aperture
correction with V=0.03 mag, B=0.05 mag, U=0.05 mag, UVW1=0.17 mag, 
UVM2($\lambda_c=2170$\AA)=0.15 mag, 
and UVW2=0.15 mag. All values plotted and listed in this paper take these
corrections into account. The data, however, are not corrected for Galactic
reddening, which is $E_{\rm B-V}$=0.050 mag \citep{schlegel98} in the direction
of the burst.

GRB~060729 was also observed by \xmm\   
\citep{jansen01} for a total of 61 ks
\citep{schartel06, campana06}. 
\xmm\ started observing the afterglow of GRB\,060729 
 on 2006 July 30 07:41 (44.9 ks after the trigger)
to 2006 July 31 01:04 UT (107.5 ks after the trigger). 
In the European Photon Imaging Camera
(EPIC) pn \citep{strueder01} the
total observing time was 59.6 ks using the medium light blocking filter.
However, due to high particle background during part of the observation
 only 42.3 ks were used.
 The observations in the EPIC MOS \citep{tur01} were for a total
observing time of 61.2 ks. The MOS1 was using the medium filter while the MOS2
observations were performed with the thin filter. The total observing time in
the Reflection Grating Spectrometers \citep[RGS; ][]{denherder01} was 61.5 ks. In the
Optical Monitor \citep[OM; ][]{mason01} the afterglow was observed for 8.3 ks in
U, 5$\times$ 4 ks in UVW1, 3$\times$ 4 ks in UVM2, and 2$\times$ 4 ks with the
optical grism. Note, that the OM UVW1 and UVM2 filters are slightly different compared
with the UVOT UV filters. The \xmm\ data were reduced with the latest SAS version
 {\it xmmsas\_20060628\_1801-7.0.0}.

\section{\label{results} Data Analysis}

\subsection{\label{position} Position of the Afterglow}
The position of the afterglow measured from the UVOT UVW1 coadded image is \\
RA (J2000) = $06^{\rm h} 21^{\rm m} 31\fs86$, \\
Dec (J2000) = $-62^{\circ} 22' 12 \farcs 5$ \\
with a 1$^{''}$ error. This position is consistent with the initial analysis of the White
and V filter analysis \citep{immler06}. The UVW1 position is $1\farcs 1$ away from the 
X-ray position 
RA (J2000) = $06^{\rm h} 21^{\rm m} 31\fs75$,
Dec (J2000) = $-62^{\circ} 22' 13 \farcs 3$
(with a $3\farcs 5$ 90\% confidence error), which was measured 
for the XRT PC mode data of segment 001 using the new
teldef file {\it swx20060402v001.teldef} as described in \citet{burrows06}.
This position deviates by $3\farcs 2$ from the
refined position given in \citet{grupe06a}. The most likely reason for this difference
is that for the X-ray 
position given in \citet{grupe06a} only the PC mode data of the
first orbit were used. This is due to the burst during the first orbit being placed on
one of the bad columns on the XRT CCD which makes the determination of a position
difficult. Figure\,\ref{grb060729_w1_image} displays the UVW1 image of the field of
GRB\,060729. The circle in the upper right inserted image is the $3 \farcs 5$ XRT error
radius of the X-ray position given above.

\subsection{BAT data}

\begin{figure*}
%\epsscale{1.5}
\epsscale{1.0}
\plotone{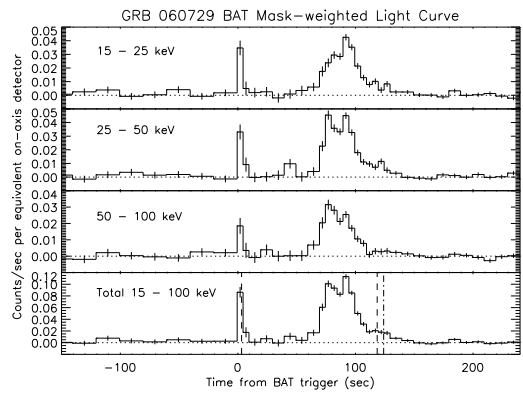}
\caption{\label{grb060729_bat_lc} \swift~BAT light curves in the 15--25 keV band
(top), 25--50 keV (second panel), 50--100 keV (second lower panel), and in the
15--100 keV band (bottom). The vertical dashed
line in the 15--100 keV plot marks the
time for which $T_{90}$ was calculated and the dotted-dashed line
the beginning of the XRT observations at 124s after the burst. 
}
\end{figure*}

\begin{figure*}
%\epsscale{1.0}
\epsscale{0.7}
\plotone{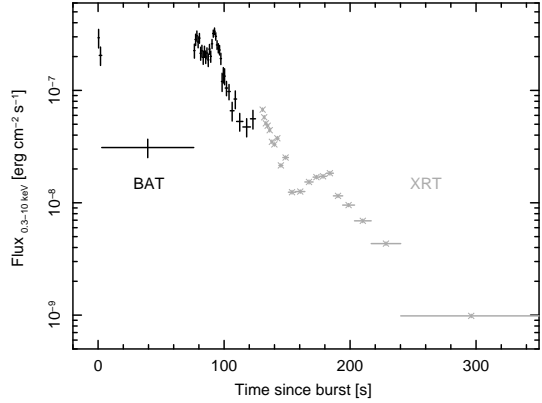}
\caption{\label{grb060729_bat_xrt_lc} Combined \swift~BAT and XRT WT light curves. The
Figure clearly shows that the XRT was starting observing GRB 060729 during the fourth
peak seen in the BAT. 
}
\end{figure*}

Figure\,\ref{grb060729_bat_lc} displays the BAT light curves in the 15--25 keV,
25--50 keV, 50--100 keV, and 15--
100 keV bands (top to bottom) with $T_0$= 2006 July 29 19:12:29 UT (Spacecraft
clock 175893150.592). 
GRB\,060729 had a $T_{90}$=115s \citep{parsons06}. Partly the $T_{90}$ is so
long because the trigger was on the pre-cursor.
After the initial first peak (pre-cursor)
which was detected by the BAT and which triggered the
observation the burst drops back down to the background level.
However, two giant peaks are observed at about 60s after the trigger, 
of which the first one is harder than the second.
There is a third peak which peaks about 120s after the trigger. This is the peak
of which we see the end of the decay in the XRT (Figure\
\ref{grb060729_bat_xrt_lc}). 
For the spectral analysis the BAT data were divided into 5 bins as listed in
Table\,\ref{bat_spec}. The first peak is the initial peak the BAT triggered on
GRB\,060729. As shown in Table\,\ref{bat_spec}, the following two peaks that occur between
70 and 124s after the burst
are by a
factor of 3 stronger than the initial peak. The two last peaks 
(124-190s after the burst, XRT flare 1 and 2 in Table\,\ref{bat_spec}) 
are also
observed simultaneously in the XRT. These data will be discussed in the XRT
section.

Table\,\ref{bat_spec} lists the results of the spectral analysis of the 5 peaks. All
spectra were fitted by a single power law model. The initial peak has a hard spectrum with
a 15--150 keV energy spectra
 index $\beta_{\rm 15-150 keV}$=1.05$^{+0.42}_{-0.32}$. 
The two strong peaks
between 70--124s after the burst show an interesting spectral behavior. While the first of
these peaks (70-88s after the burst) has a rather hard spectral slope with 
$\beta_{\rm 15-150 keV}$=0.59\plm0.11, the second of these peaks (88-124s)
was softer with $\beta_{\rm 15-150 keV}$=0.90\plm0.11. 
The total fluence in the observed 15-150 keV band is 2.7$\times 10^{-6}$ ergs cm$^{-2}$
\citep{parsons06} and 7.2$\times 10^{-6}$ ergs cm$^{-2}$ and 
1.7$\times 10^{-5}$ ergs cm$^{-2}$ 
in the rest-frame 1 keV -- 1 MeV and 
1 keV -- 10 MeV bands, respectively, adding all BAT spectra together
(Table\,\ref{bat_spec}) and assuming the same power law spectrum as in the 15-150 keV
band without any break.
 With the redshift z=0.54 this converts into an isotropic energy in the 
 rest-frame 1 keV - 1 MeV and 
 1 keV -- 10 MeV
 band of $E_{\rm iso}=6.7\times 10^{51}$ ergs and
 $E_{\rm iso}=1.6\times 10^{52}$ ergs, respectively. Because we lack observations of the break
 energy $E_{\rm break}$ and the $\gamma$-ray spectrum at higher energies by Konus-Wind the 1 keV -- 10
 MeV band $E_{\rm iso}$ value is an upper limit of the true isotropic energy.
 
\subsection{X-ray data}

\subsubsection{Temporal analysis}

Figure\,\ref{grb060729_bat_xrt_lc} shows the combined BAT and XRT light curve.
The light curve clearly shows that XRT began observing the GRB at the beginning 
of the fourth peak seen in the BAT.  The combined BAT + XRT light curve was
constructed as described in \citet{obrien06}. Due to the dramatic spectral
change within the three minutes of the WT observation, we applied an ECF for each
individual bin assuming power law model corrected for absorption 
with the parameters listed in Table\,\ref{wt_xspec}.
However, for the BAT data we applied only one ECF that reflects
the main spectrum.

\begin{figure*}
%\epsscale{1.0}
\epsscale{1.2}
\plotone{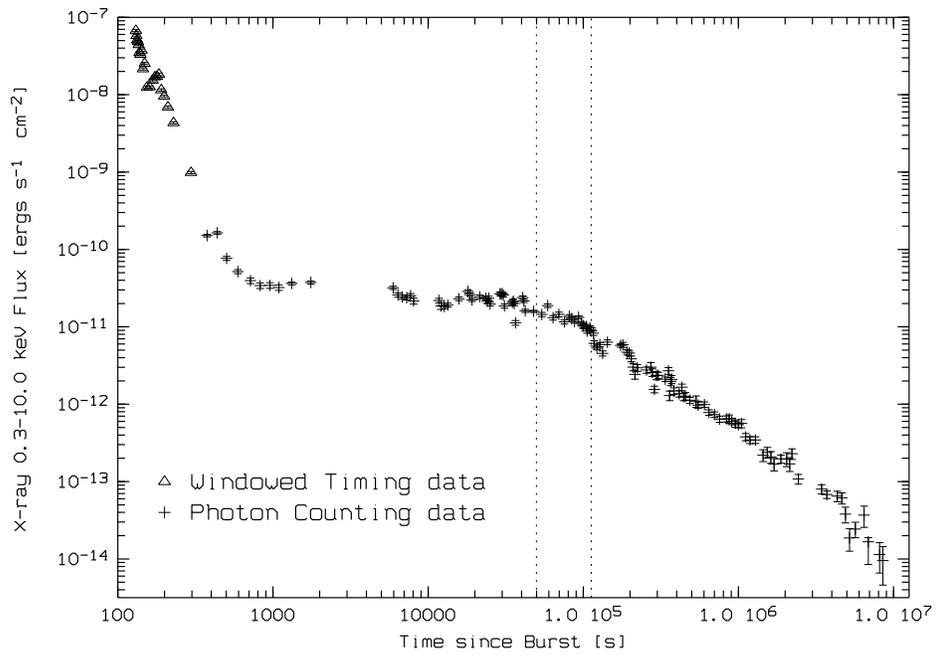}
\caption{\label{grb060729_xrt_lc} \swift~XRT light curve of the Windowed Timing
(triangles) and photon counting (crosses). The downward arrow marks the 
3$\sigma$ upper limit at
the end of the \swift\ observations. This upper limit contains a total exposure time
of 63.5 ks obtained between 2006 December 8 and December 27.
The dotted vertical lines mark the
start and end times of the \xmm\ observation.  
}
\end{figure*}

Figure\,\ref{grb060729_xrt_lc} displays the \swift~XRT light curve, with
WT mode data as triangles and PC mode data as crosses. The vertical dashed lines
in the figure mark the start and end times of the \xmm\ observations. The general behavior
of the light curve can be described as follows: after the initial steep decay with a decay
slope  $\alpha_{\rm 1}$ = 5.11\plm0.22 the light curve flattens at
$T_{\rm break, 1}$ = 530s\plm25s with a decay slope
$\alpha_{\rm 2}$ = 0.14\plm0.02. At
$T_{\rm break, 2}$ = 56.8 ks\plm10ks the light curve of the afterglow breaks again and
continues decaying with a decay slope
$\alpha_{\rm 3}$ = 1.29\plm0.03.
The definitions of the decay slopes follow the descriptions given in
\citet{nousek06} and \citet{zhang06}.
 We do not detect a jet break even \days days after the burst. 
 The last 3$\sigma$ detection of
 the X-ray afterglow was obtained between 2006 November 21 to December 01 with a total
 exposure time of 69.9 ks. The afterglow was still observed by \swift\ until December
 27 for a total of 63.5 ks. However these observations were interrupted by several new
 bursts and at the end only a 3$\sigma$ upper limit of 2.1$\times 10^{-14}$ ergs
 s$^{-1}$ cm$^{-2}$ could be obtained. It was dropped from the \swift\ schedule after
 2006 December 27 because it was not detectable anymore with the XRT within a
 reasonable amount of observing time.

\subsubsection{Spectral Analysis}

\subsubsubsection{Dramatic spectral change at the beginning}

In order to examine the spectral behavior in more detail,
source and background spectra were created for each bin, except for the first 10
bins ($T-T_0$=130--150 s). The WT mode data were divided into 21 bins with 1000 source photons in each bin.
Because of the high count rate at the beginning of the observations we applied
the method as described in \citet{romano06} to avoid the effects of pileup.
 However, this procedure reduced significantly the number of
source photons in each single spectrum. We, therefore, combined two 
bins into one for bins 1+2, 3+4, 5+6, 7+8, and 9+10 
 to increase the signal to noise ratio. 

\begin{figure*}
%\epsscale{1.5}
\epsscale{1.5}
\plotone{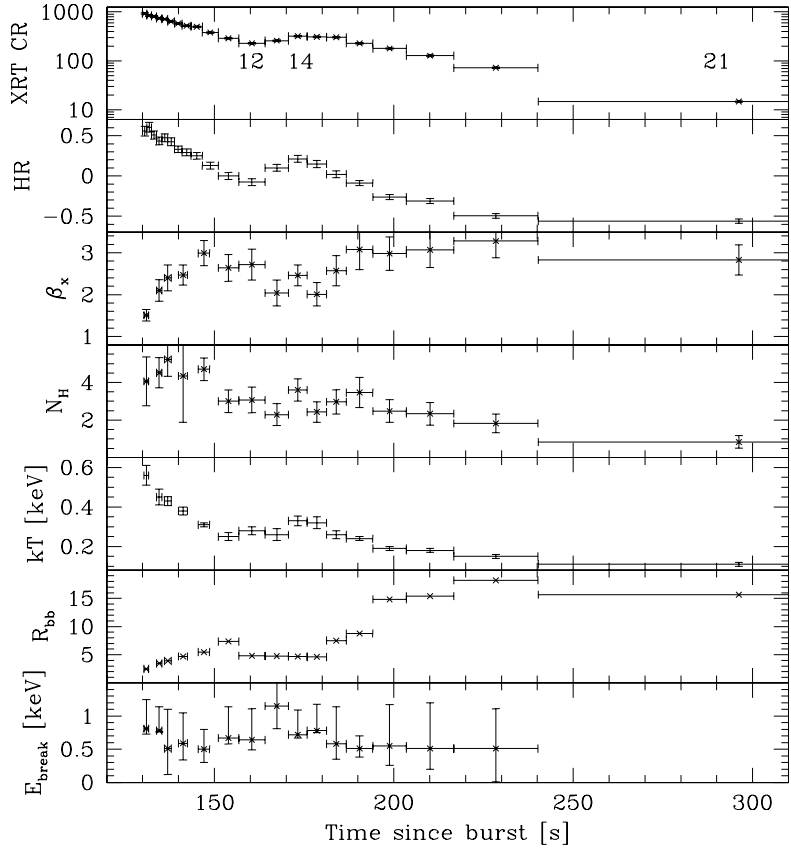}
\caption{\label{grb060729_wt_lc} \swift~XRT Windowed Timing mode light curve.
The panels display from the top to the bottom the XRT count rate (in
units of counts s$^{-1}$),
the hardness ratio
(see text for definition), the X-ray spectral slope \bx\ of a single power law fit,
the free fit column density $N_H$ in units of $10^{21}$ cm$^{-2}$,
the blackbody temperature kT, the blackbody radius $R_{\rm bb}$ (in
units of $10^{12}$ cm), 
and the cut off energy $E_{\rm break}$ of a 
power law with exponential cutoff. 
All these fit parameters are listed in Table\ \ref{wt_xspec}.
The numbers in the top panel mark the bins that
were used for the spectra shown in Figure\ \ref{grb060729_wt_spectra}.
}
\end{figure*}

Each of the 16 spectra were fitted by an absorbed single power law, blackbody
plus power law, and a power law with exponential cutoff model. 
The results of these spectral fits are listed in Table\ \ref{wt_xspec}. 
Figure\,\ref{grb060729_wt_lc} shows plots of the WT mode data; from top to
bottom
the count rate, hardness ratio, 
the X-ray spectral slope \bx\ from an absorbed single powerlaw fit with a free fit
absorption column density $N_H$, 
blackbody temperature kT (in
keV) from the blackbody plus power law fit, the blackbody 
radius\footnote{The blackbody radii were derived for each bin from the relation
$L=4\pi R_{\rm bb}^2 \sigma T^4$, where $\sigma$ is the Stefan-Boltzmann-constant} 
$R_{\rm bb}$
and the break energy $E_{\rm break}$ 
of a power law with exponential cutoff
model.   The hardness ratio changes from HR=0.6 at the beginning of the observation
to HR=$-$0.56 at the end, indicating a dramatic evolution in the X-ray
spectrum within 2 minutes of observing time which translates into 1.3
 minutes in the
rest frame. 

While the spectra during the first 20 s of the WT mode observation are
well-fitted by a single power law model, after about 150s after the burst the
data are better fit by a blackbody plus power law spectrum.
For the absorbed power law fits, all parameters were left free. We found that
while the spectra at the beginning of the observation were rather hard with
\bx=1.5 and \nh=$4\times 10^{21}$ cm$^{-2}$, the spectra became very soft with
\bx$\sim$3.0 and \nh=$1\times 10^{21}$ cm$^{-2}$. Note that the absorption
column density \nh\ decreases while the energy spectral index \bx\ becomes
steeper - the opposite than what is expected if the spectral slope and the
absorption column density were just linked in the fitting program. 
However, note that especially during
the later bins, the spectra are not well fit by a single power law and do require
more complicated models. 

For the blackbody
plus power law model, the absorption parameter was fixed to the Galactic value
\citep[4.82$\times 10^{20}$ cm$^{-2}$][]{dic90}
and the hard energy spectral slope to \bx=1.0. The blackbody temperature changes
dramatically from $kT$=0.56 keV at the beginning of the XRT WT observation 
to 0.11 keV at the end, accompanied by an increase of the blackbody radius from
2.5$\times 10^{12}$ cm to 16$\times 10^{12}$ cm. Fitting the data with an absorbed blackbody plus power law model
with the absorption column density at z=0 set to the Galactic value and at z=0.54 set to
$1\times 10^{21}$ cm$^{-2}$ (see the discussion about the \xmm\ spectral analysis)
results in similar values for the temperature. The only differences are that the
temperatures tend to be lower by 40 eV and the normalizations are higher. 

The prompt emission of GRBs is often fitted by a Band function \citep{band93}.
We also tried a power law model with exponential cutoff, which is a surrogate
for the Band model, which has the advantage of using fewer parameters than the
Band model. In order to obtain better constraints we fixed the absorption column
to the Galactic value. As listed in Table\,\ref{wt_xspec}, typically the power
law model with exponential cutoff does not show improvements over the single
power law or the blackbody plus power law models.

\begin{figure*}
\epsscale{1.0}
%\epsscale{0.6}
\plotone{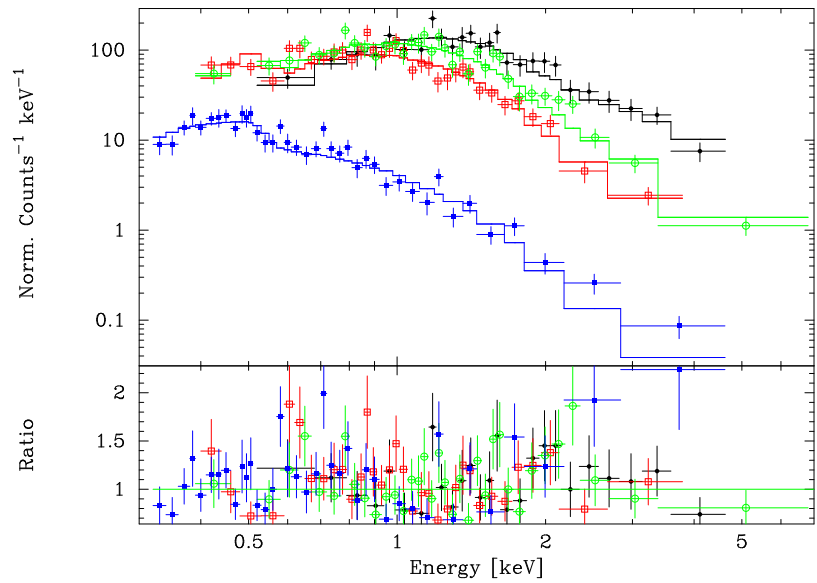}
\caption{\label{grb060729_wt_spectra} \swift~XRT WT spectra of bins 
1+2 (black filled circles),
12 (red open squares), 14 (green open circles), and 21 (blue filled squares) 
fitted by absorbed single power laws as
given in Table\,\ref{wt_xspec}.
The numbers refer to the bins as shown in
Figure\ \ref{grb060729_wt_lc}.
}
\end{figure*}

The change in the X-ray spectra is also displayed in 
Figure\,\ref{grb060729_wt_spectra} which shows the spectra of bins \# 1, 12, 14, and 21. 
Bin 12 is the bin before the small flare at 170s after the burst and bin 14 is the peak of
that flare.  

%Figure\,\ref{grb060729_wt_corr} displays the dependence of the energy spectral slope
%\bx, the blackbody temperature kT, and the cutoff energy $E_{\rm break}$ with the
%count rate. All these plots suggest strong correlation with the count rate.  All
%correlation of \bx, kT, and $E_{\rm break}$ show very strong correlation with the
%count rate. In all cases the probability of a random result is P$>$0.0001.

\subsubsubsection{Later PC mode data}

All PC mode data can be fitted by a power law model with a
energy spectral slope \bx=1.2 and an absorption column density of about 1.5$\times
10^{21}$ cm$^{-2}$. This absorption column density is significantly above the Galactic
value . The intrinsic absorption column
density at the redshift z=0.54 is 1.9\plm0.4$\times 10^{21}$ cm$^{-2}$. 
Table\,\ref{xmm_xrt_xspec} lists the XRT PC mode observations at 20--40 ks after the 
burst and at 200 ks after the burst, so before
and after the XMM observation. The fits to these data suggest no significant spectral 
variability before and after the break in the
X-ray light curve around 60 ks after the burst.

\subsubsubsection{The \xmm\ observations}

\begin{figure*}
\epsscale{1.0}
%\epsscale{0.6}
\plotone{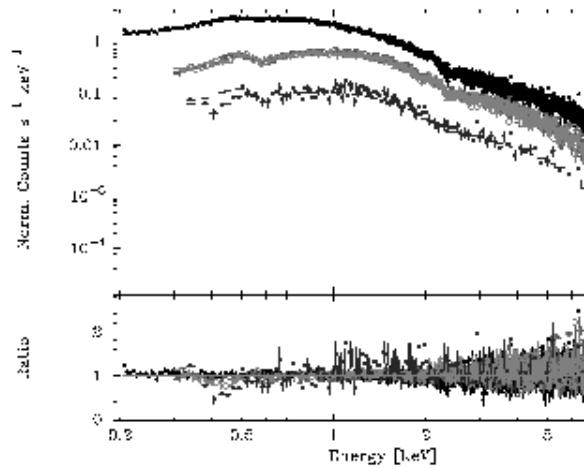}
\caption{\label{grb060729_xmm_xrt_spectra} 
Absorbed power law model fits with \nh=8.57$\times 10^{20}$ cm$^{-2}$ and
\bx=1.08, as listed in Table\,\ref{xmm_xrt_xspec},
to the 
\xmm\ EPIC pn (black filled circles),
MOS1 (red open squares), MOS2 (green open circles), and \swift~XRT PC 
mode(blue filled squares) spectra. 
The \swift~XRT data were selected between 44.9 ks to 107.5 ks after 
the burst, simultaneous with the \xmm\ observation.
}
\end{figure*}

The combined spectra of the \xmm~EPIC pn and MOS and \swift~XRT data are shown
in Figure\,\ref{grb060729_xmm_xrt_spectra}. The XRT data were selected
between 44900 s and 107500 s after the burst. The details of the spectral
fits to these data are summarized in Table\,\ref{xmm_xrt_xspec}. At these late times, the
X-ray spectra were
well fitted by absorbed single power law models. 
However, as a check the spectra were fitted also by a
blackbody plus power law model, but at these late times the power law component dominates
the spectra. Therefore, we will only discuss the absorbed power law
model fits as listed in Table\,\ref{xmm_xrt_xspec}.

The obvious 
difference between the  \swift~XRT and \xmm~pn and MOS data is the 
much higher value of the absorption column density. From the free fit absorption column
density at z=0  we measured an absorption
column density \nh=15.7$\times 10^{20}$ cm$^{-2}$ in the \swift~XRT data. This value is
about twice as high as what is measured from the \xmm~EPIC pn and MOS spectra. The EPIC pn
is well-calibrated to energies below 0.2 keV \citep[e.g. ][]{haberl03}. 
 We also applied the {\it gain fit} model
within  {\it XSPEC}, but it did not remove the discrepancy.
This discrepancy maybe due to problems with the \swift~XRT bias maps during the
time period 2006 July 21 and 2006 August 3. This bias map problem causes an
offset in the gain and therefore compromises the spectral analysis of
 \swift~XRT PC mode data
during that time period. However, this gain shift does not affect the early 
\swift~XRT WT mode data. 
Due to the better response of the EPIC pn at lower energies we
consider the absorption column densities measured by the EPIC pn the most reliable.
With the redshift of the burst at z=0.54 we can also use the X-ray spectra to determine
the intrinsic absorption column density at the location of the afterglow. The intrinsic
column densities of all fits are in the order of $1\times 10^{21}$ cm$^{-2}$, except for
the \swift~XRT data which again show an absorption column density about twice as high.
As we will show later in section\,\ref{sed} the absorption column density of $1\times
10^{21}$ cm$^{-2}$ is in good agreement to what can be derived from the spectral energy
distribution of the afterglow.

In addition to the EPIC pn and MOS data we also analyzed the 2 RGS spectra. We found that
the analysis of the RGS continuum spectra agrees within the errors with the pn and MOS 
data. We did not find any obvious emission or absorption features in the RGS spectra.

\subsection{\label{uvot_analysis} UV/Optical data analysis}

\begin{figure*}
\epsscale{1.8}
%\epsscale{1.0}
\plotone{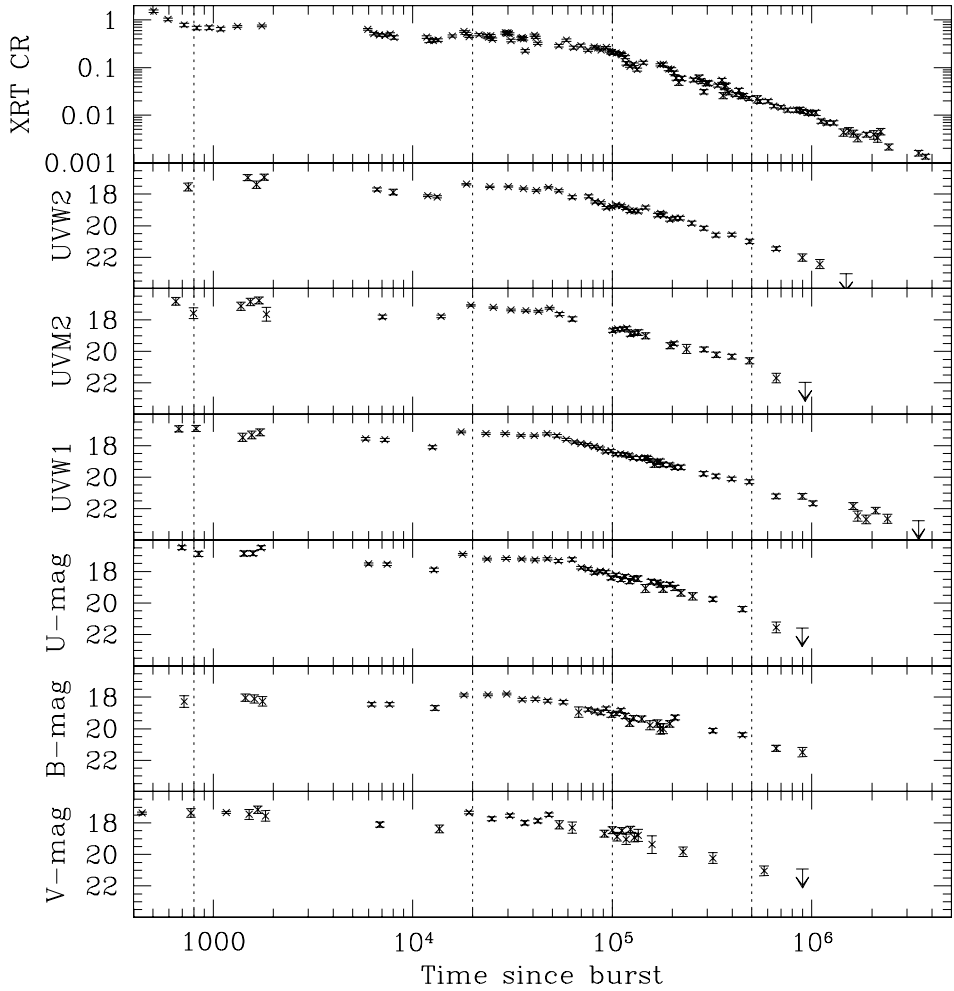}
\caption{\label{grb060729_xrt_uvot_lc} \swift~UVOT and XRT PC 
mode light curves, with the XRT light curve on the top and the UVOT
light curves starting with UVW2 on the top and V on the bottom. The
downward arrows mark the 3$\sigma$ upper limits as listed in
Table\,\ref{uvot_lc}. The vertical lines mark the times when the spectral energy
distributions of the afterglow were determined as shown in Figure\,\ref{grb060729_sed}.
}
\end{figure*}

The magnitudes resulting from the UVOT data analysis are 
listed in Table\,\ref{uvot_lc}.
Figure\,\ref{grb060729_xrt_uvot_lc} shows the results of the UVOT photometry in
comparison with the XRT. UVOT was able to follow the afterglow in all six
filters up to 9 days after the burst. In UVW1 the afterglow was followed up
31 days after the burst which translates into 20 days in the rest-frame.
This is one of 
the longest intervals \swift's UVOT has ever detected an afterglow in the 
optical/UV. Only GRBs 060218 \citep{campana06b} and 060614 \citep{mangano07}
were detected at slightly later observed
times than GRB\,060729. 

In all bands,
XRT as well as in all 6 UVOT filters, a significant break occurs in the light 
curve.
Table\,\ref{uvot_lc_slope} lists the decay slopes $\alpha_{\rm 2}$ and 
$\alpha_{\rm 3}$
before and after the break time $T_{\rm break}$. Within the errors all break
times seem to occur at about 50 ks after the burst (33 ks in the rest-frame),
with the earliest break in B at about 30ks and the later breaks at shorter
wavelengths. However, considering the uncertainties in the decay slopes, this is
all consistent with an achromatic break. Note that due to a rebrightening of the afterglow at about 20 ks seen in X-rays and all 6
UVOT filters the determination of $\alpha_2$ is rather uncertain.
 In B the afterglow decays the
slowest with $\alpha_3$=0.98. The decay slopes at shorter wavelengths are steeper
with $\alpha_3 \approx$=1.3. Note that the flatter slope in the B-Filter is caused by a 
re-brightening at about 200 ks that is not seen in the other filters. 
By limiting the 
analysis to data only up to 200ks the decay slope is $\alpha_3$=1.17\plm0.16 which is
consistent with the decay slopes seen in the other filters.   
The rebrightening in B at about 200 ks after the burst seems to be real. We
checked for any strong variability in the background but could not detected any
strong background variation at that time. The decay slopes $\alpha_2$ and
$\alpha_3$ are consistent with the decay slopes reported by \citet{quimby06b}
and \citet{cobb06}. 

\begin{figure*}
\epsscale{2.0}
%\epsscale{1.0}
\plottwo{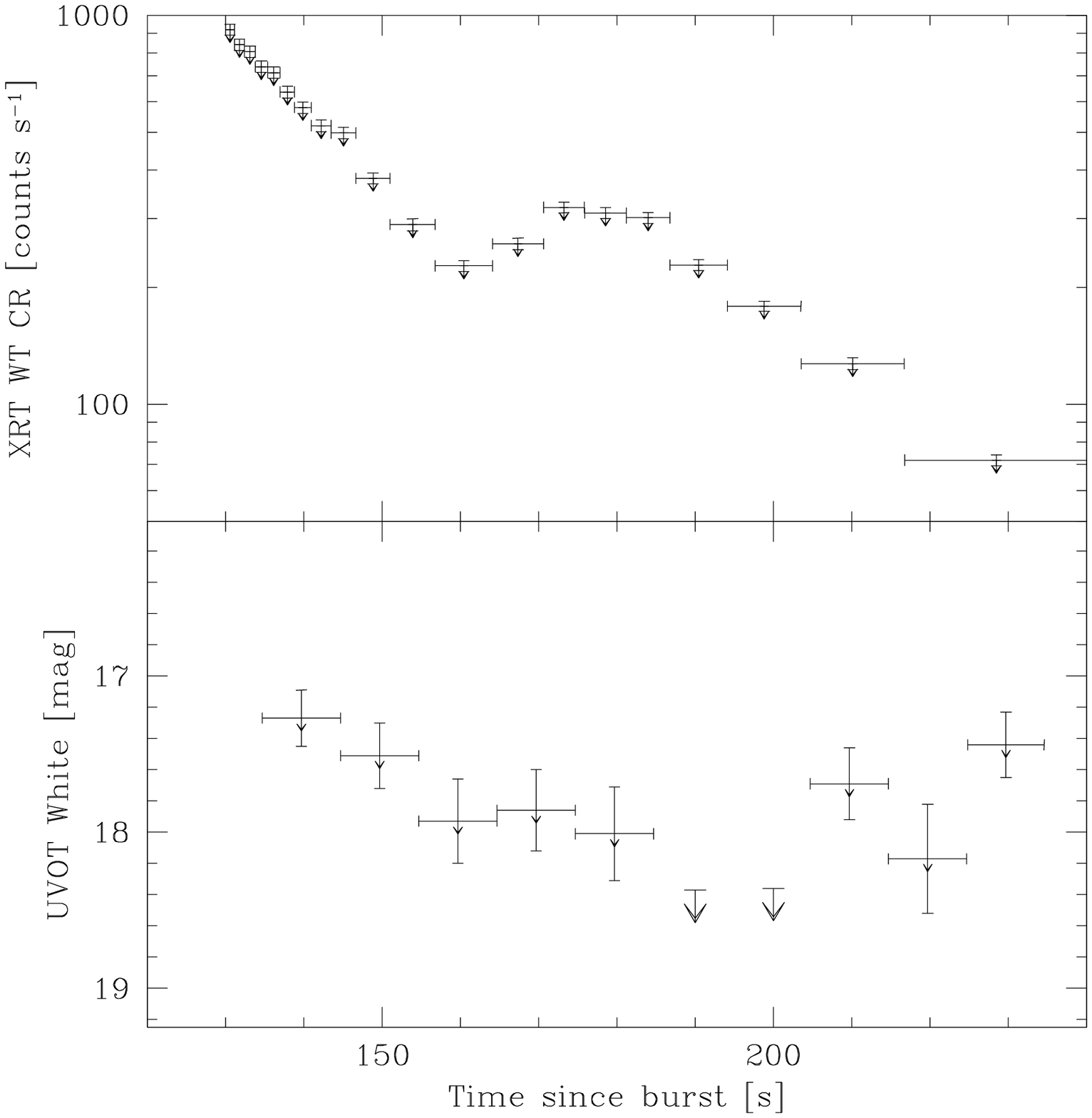}{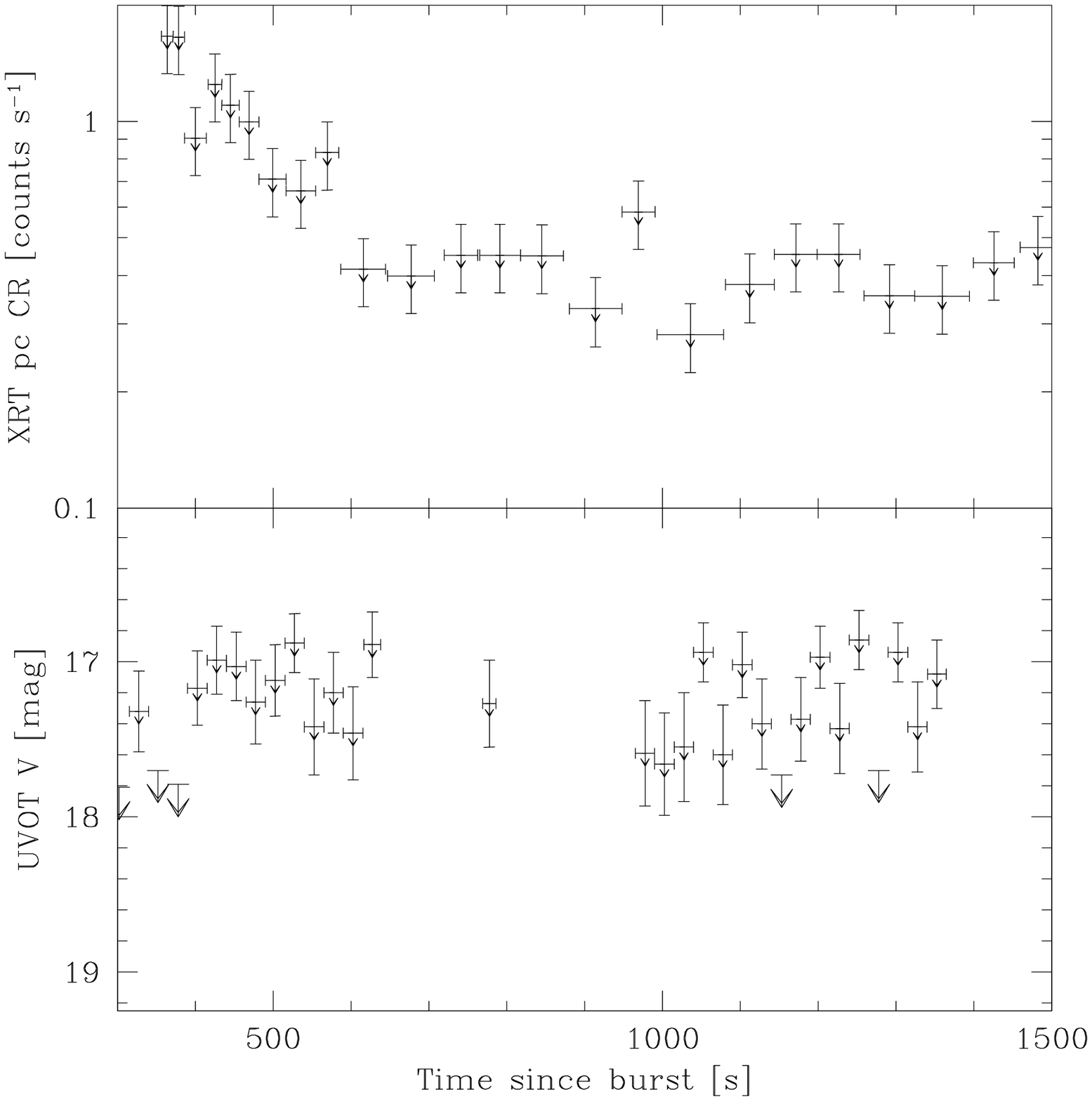}
\caption{\label{grb060729_xrt_uvot_early} \swift~UVOT White and V
and XRT WT PC mode early time light curves of the first orbit. The left panel shows the
UVOT White and the XRT WT mode data, and the right panel shows UVOT V and XRT PC mode data.
The binning in the PC mode data is 25 photons  per bin. 
The UVOT White data are binned in 10 s
intervals and the V data are binned in 25 s intervals.
}
\end{figure*}

Figure\,\ref{grb060729_xrt_uvot_early} displays the UVOT White and V 
and XRT light curves of the first orbit. The UVOT data of this period are listed
in Table\,\ref{uvot_early}. The
left panel of Figure\,\ref{grb060729_xrt_uvot_early} displays the UVOT White filter event
mode and XRT WT mode data. The UVOT White filter data were grouped into 10 s bins.
The first UVOT White points show a decay similar to the XRT WT light curve. However after
these few points the UVOT White light curve flattens, which agrees with the flare seen in
the XRT data at 170s after the burst. At about 200s after the burst the afterglow starts
to become brighter in the UVOT White, while it is still decaying in X-rays.

The right panel of Figure\,\ref{grb060729_xrt_uvot_early}
shows the UVOT V event mode data and the XRT PC mode data of the first orbit. The UVOT V
data were grouped in 25 s bins and the XRT PC mode data with 25 source photons per bin.
The UVOT V light curve shows the afterglow fairly constant at about 17.5 mag while the XRT
PC mode light curves shows an initial decay until about 600s and flattens after that. 

The \xmm~Optical Monitor observations are summarized in Table\,\ref{om_data}. While the U
filter results agree well with the UVOT U data as listed in Table\,\ref{uvot_lc}, there
is a discrepancy in the UVW1 and UVM2 filters. There maybe two explanations for this
discrepancy: 1) The OM and UVOT UV filter sets have different filter transmission, 2)
OM suffers from significantly higher level of scattered light than
the UVOT, and third the
extraction radius of the automated OM software is 12$^{''}$ which is too large for an
accurate analysis of the UV data due to the bright star (see UVOT section). 
The brightening
of the afterglow in the last OM UVM2 observation at 101 ks after the
burst is most
likely due to a bad subtraction of the background within the
automatic OM data reduction software.

\subsection{\label{sed} Spectral Energy Distribution}

\begin{figure*}
\epsscale{1.8}
%\epsscale{1.0}
\plotone{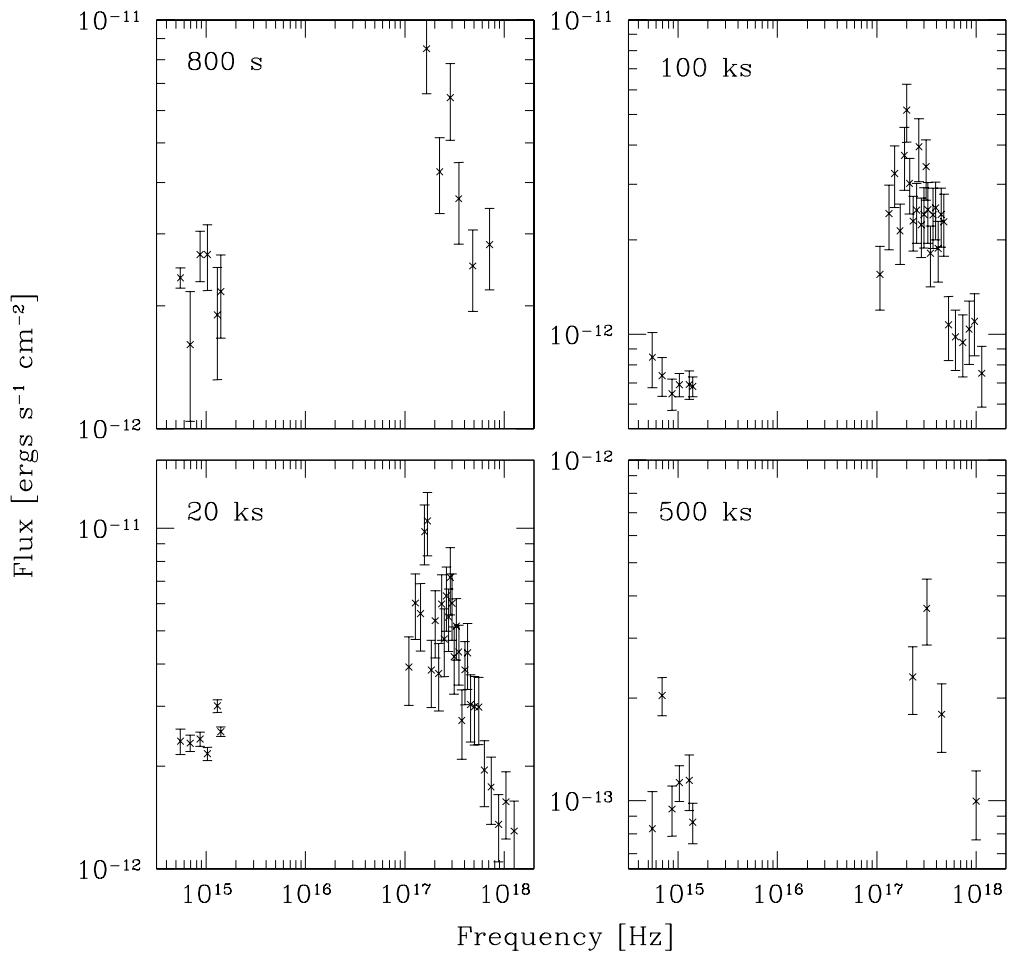}
\caption{\label{grb060729_sed} Spectral energy distributions of the afterglow of GRB\,060729 at 800s (upper left),
20 ks (lower left), 100ks (upper right), and 500 ks (lower right). The UVOT
photometry data are corrected for Galactic reddening \citep[$E_{\rm B-V}$=0.050][]{schlegel98}.
}
\end{figure*}

As shown in section\,\ref{uvot_analysis} the long-term light curves in all 6 UVOT
filters and in X-rays do follow the same decay
slope with similar break times at about 50 ks
after the burst. In order to check this we determined spectral energy distributions
(SEDs) of
the afterglow at 800s, 20 ks, 100 ks, and 500 ks, {\bf with exposure times
in the XRT of 400s, 1.8 ks, 3.3 ks, and 4.8 ks, respectively.}
 Figure\,\ref{grb060729_sed} displays
these four SEDs. These times are also marked in Figure\,\ref{grb060729_xrt_uvot_lc}.
These times were picked to represent the SEDs of the earliest and latest time possible
when the afterglow was detected in all 6 UVOT filters and shortly before and after the
break. All fluxes in all 6 UVOT filters and the XRT were calculated according
to the light curves. 
There seem to be no obvious changes in the SEDs, besides the changes in the
fluxes, over time. Another measure of any changes in the SEDs over time is
the optical/UV to X-ray spectral slope or X-ray loudness\footnote{The
X-ray loudness is defined by \citet{tananbaum79} as \aox=--0.384
log($f_{\rm 2keV}/f_{2500\AA}$).} \aox.  For the
afterglow of GRB\,060729 we measured rest-frame \aox~of 0.85\plm0.10 at 800s, 0.84\plm0.05
at 20 ks, 0.74\plm0.07 at 100 ks, and 0.77\plm0.10 at 500 ks after the burst. Within the
errors these values are consistent and do not suggest any changes between the optical/UV
and X-ray parts of the SED over time after the first orbit. However, note that
during the first $\approx$400s of data the SED changes dramatically, because the
X-ray flux decays very fast ($\alpha_{x,1}$=5.1 while the White and V data
suggest that the optical afterglow is constant.   
A single power law spectrum between optical/UV and X-ray energies a \aox=1.1 is
expected according to the fits to the X-ray data (Table\,\ref{xmm_xrt_xspec}). 
This assumption of a single power law spectrum between the optical and X-rays is justified
given that the optical has the same temporal behavior as the X-rays and that the
optical and X-rays are both above the cooling frequency.
The
difference between the expected \aox=1.1
and the measured \aox$\approx$0.8 values suggests intrinsic reddening
at the location of the afterglow. Based on the absorption corrected rest-frame 2 keV flux
density we can calculate the expected flux density at rest-frame 2500\AA. We calculated a
reddening of 1.7 mag at rest-frame 2500\AA\ which corresponds to an $E_{\rm B-V}$=0.34
mag. Applying the relation given by \citet{diplas94}\footnote{The \citet{diplas94}
relation is: $N_{\rm H} = 4.93\times10^{21}\times E_{\rm B-V}$ cm$^{-2}$.}
we calculated an intrinsic
column density $N_{\rm H, intr}=1.7\times 10^{21}$ cm$^{-2}$. 
Considering that this is a rough estimate, this absorption column density agrees
quite well with that measured from the \xmm~EPIC pn spectrum,
$N_{\rm H,intr} = 0.85\pm0.02 \times 10^{21}$ cm$^{-2}$
(Table\,\ref{xmm_xrt_xspec}).

\section{\label{discuss} Discussion}

The afterglow of 
GRB\,060729 has been detected in X-rays by the \swift-XRT
longer than any other 
\swift-detected burst, up to 125 days after the burst.
Finally by the end of December we had to give up on observing this burst by \swift\ 
because it became too faint to be detectable anymore in the XRT detector.
Even though the afterglow was dropped from the \swift\ observing schedule after 2006
December 27, we are still planning to obtain more observations with larger
observatories such as \chandra, \xmm, and {\it Suzaku}.

\subsection{Light curves}

The X-ray and UV/Optical light curves are remarkably similar. 
Not only that their
decay slopes and break times are in good agreement (except for the B light
curve), but they also seem to be synchronized
 during rebrightening phases, which can
be seen best in the UVW2 light curve. In particular the re-brightening at about 15
ks after the burst clearly appears to be present in all 6 UVOT filters and in
X-rays. Even though the break at around 50 ks after the burst seems to be
achromatic, we do not consider this to be a jet break. The post-break 
decay slope
$\alpha_{\rm X,3}$=1.29 is to shallow to be a jet break. 
If we interpret this slope as the post jet break slope, we would get $p=1.29$,
which is much flatter than the electron distribution index  predicted by shock
acceleration theory (usually $p\ga2$). Furthermore,  if  $p=1.29$, one would
expect the spectral slope $\beta_x=p/2=0.65$ or $\beta_x=(p-1)/2=0.15$ for a
cooling frequency below or above the X-ray range respectively, which are
obviously contrary to the observed spectral slope of \bx=1.1
Most likely we have not seen the jet break in the afterglow of 
GRB 060729 because the afterglow has not been followed long enough as the studies
by \citet{willingale07, sato07} suggest.

One of our main results is that 
the light curve of the afterglow 
does not yet show a jet break at 
\days days after the burst  (81 days in the rest frame).
This is the longest period a GRB afterglow was ever followed and detected in
X-rays, except for
GRB 030329 which was followed  
258 days after the burst by \xmm\ \citep{tiengo04}. According to the relation
given in \citet{willingale07} we would have expected to see a jet break at
about 5.5$\times 10^6$ s after the burst.
With an isotropic energy in the rest-frame 1 keV-10 MeV $E_{\rm
iso}=1.6 \times 10^{52}$ ergs and the relations given by \citet{sari99} and
\citet{frail01}, and the non-detection of a jet break up to 125 days after the
burst,
we derive that the opening angle of the jet has to be larger than
28$^{\circ}$, assuming a particle density $n=0.1$cm$^{-3}$ and an efficiency
$\eta_{\gamma}$=0.2.
With $\beta_{\rm X}$=1.0 and $\alpha_{\rm X}$=1.29 according to Table 2 in \citet{zhang06}
we estimated an electron index slope p=2.3 for the ISM or wind case with 
$ \nu>\max(\nu_m,\nu_c)$.

The afterglow of GRB\,060729 is not only remarkable for its long follow-up observations in
X-rays but also for its relatively late break time between the flat decay phase and the
steepening phase \citep[phases 2 and 3 according to ][]{zhang06, nousek06,willingale07} 
at about 53 ks after the burst, which converts to 35 ks in the rest frame. 
 Typically the break between
 phases 2 and 3
occurs around 10 ks after the burst \citep{nousek06, willingale07}. 
The late time break at 35 ks
after the burst (rest frame) requires a substantial ongoing injection of energy into the
afterglow. In a matter of fact, when we observed this afterglow with Swift e at
first did not detect a break in the light curve until about 2 days after the
burst due to the lumpiness of this plateau phase. Also note that
\citet{willingale07} list the break time in the X-ray light curve at 120 ks
after the burst.

The relatively long time of this very flat decay
($\alpha_{x,2}=0.14\pm0.02$) phase implies larger energy injection
during the early time in this burst than in other bursts.
This energy injection could
be the result of a refreshed shock \citep{rees98} or continuous
energy input from the central engine.
 Assume
the energy injection has the form $L(t)\propto t^{-q}$, the light
curve decay rate is $\frac{(2p-4)+(p+2)q}{4}$ for
$\nu_X>\max({\nu_m, \nu_c})$ \citep{zhang06}.
So we get
$q\simeq0$, and the total  energy is increased by a factor of
$(T_{break,2}/T_{break,1})^{1-q}\sim100$ during this energy
injection phase, where $T_{break,1}$ and $T_{break,2}$ are the
first and second break time of the x-ray light curve respectively.
Such a large energy increase factor is the highest among the
\swift-detected GRBs \citep{nousek06}. 
This may be part of the
reason that we see a bright X-ray afterglow for a very long time,
besides that no jet break occurs before \days days after the burst.
The plateau phase of the X-ray afterglow of GRB\,060729 is one of the longest
ever observed by \swift. The total fluence in the 0.3-10.0 keV band of this
plateau is about 1$\times 10^{-6}$ ergs cm$^{-2}$, which is about 1/3 of the
total 15-150 keV fluence of the prompt emission.
The $q$ value ($q=0$) inferred for this burst implies a pulsar
type (i.e. magnetic dipole radiation) energy injection \citep[e.g. ][]{dai98,
zhang01} 

Using the X-ray luminosity $L_x=2\times10^{46}$ergs s$^{-1}$ at
$t=10~ {\rm h}$, we estimate the isotropic kinetic energy at this
time is $E_{k,iso}\sim2.5\times10^{53}(\epsilon_e/0.1)^{-1}{\rm
ergs}$ \citep{freedman01, zhang07},
where
$\epsilon_e$ is the equipartition factor of electrons in afterglow
shocks. Assuming the energy increasing factor 100 during the flat
decay phase, the isotropic kinetic energy before the energy
injection phase is only
$\sim2.5\times10^{51}(\epsilon_{e}/0.1)^{-1}{\rm ergs}$, implying a
high efficiency of gamma-ray production during the prompt phase.
Using the isotropic kinetic energy and the jet break time larger
than \days days after the burst, we get the jet opening angle larger
than $\theta_j\ga28^\circ n_{-1}^{1/8}$ \citep{frail01},
 where
$n_{-1}\equiv (n/0.1{\rm cm^{-3}})$ is the number density of the
circumburst ISM. From this, we further get the beam-corrected
kinetic energy of the burst $E_{k,j}\ga1.7\times10^{52}{\rm
ergs}(\epsilon_e/0.1)^{-1}n_{-1}^{1/4}$. Such a larger kinetic
energy than those in usual bursts may be  a direct consequence of
the unusually long energy injection phase in the early time.

Even though the X-ray and UV/optical afterglow of GRB 060729 was unusually
bright, it was rather unspectacular in the BAT 15-150 keV energy range. The
15-150 keV fluence of 2.7$\times 10^{-6}$ ergs cm$^{-2}$ \citep{parsons06} 
is rather moderate compared to
other \swift-discovered bursts. Also note, that the peak luminosity of GRB
060729 of about 3$\times 10^{50}$ ergs s$^{-1}$ is very low for a long burst
regarding the time lag between the 50-100 keV and 15-25 keV band as shown by
\citet{gehrels06}.

The observations of the X-ray and UV fields of GRB 060729 have been the 
deepest ever performed by \swift.
In X-rays we observed the field  for 1.13 Ms and for 550 ks each in UVW1 and 
 UVW2. The
UVW1 and UVW2
observations are the longest exposure that
were taken of any field in the UV by any UV
observatory. 
The results of this
study and the source identifications based on their spectral energy distributions will be
presented in a separate paper which is in preparation.

\subsection{Spectra analysis of the early time data}

During the XRT WT mode  epoch of observations, two flares are detected.
For the first X-ray flare, only the decay part is seen by the XRT. The
decay rates after the peak of the flares are as steep as $t^{-5}$,
pointing to internal central engine activity as the 
origin for the X-ray flares \citep{burrows05b, fan05, zhang06, dai06, falcone06,
wu06, wang06, lazzati06, gao06}.

%(Burrows et al. 2005b; Fan \& Wei 2005;
%Zhang et al. 2006; Dai et al. 2006; Falcone et al. 2006; Wu et al.
%2006; Wang, Li \& Meszaros 2006; Lazzati \& Perna 2006).

During 130-160 s (time bins from 1 to 11 in Table\,\ref{wt_xspec}) and 190-300 s, the X-ray
emission is undergoing a steep decay, which may result from the
high-latitude emission of the corresponding flares \citep{kumar00, liang06},
%(Kumar \& Panaitescu 2000; Liang et al. 2006), 
i.e. we see the curvature
effect of the radiation pulse. In this picture, the count rate
and the peak energy ($kT$ or $E_{\rm break}$) both decrease as the
pulse decays. This is because less and less Doppler-boosted radiation is
seen from the pulse. This accounts for  the strong correlation between
the count rate and the blackbody temperature $kT$ (Spearman rank order
correlation coefficient $R_{\rm S}$=0.96 with $T_{\rm S}$=11.8 and a
probability of P$< 10^{-4}$ of a random distribution; linear correlation
coefficient $r_{\rm l}$=0.95).
%($kT$ or $E_{\rm break}$).

The WT data of the early afterglow observations showed a dramatic change in the X-ray
spectrum within less than two minutes in the rest frame. Typically the spectra of the
bins of the WT mode data can be fitted by a single absorbed power law or a black body
plus power law model. The power law model shows that there is a decrease in the
absorption column density by a factor of 4 from the beginning of the observation at 131 s
after the burst to 300s after the burst. Such decreases in the absorption column density
have been observed before in GRB afterglows, like e.g. in GRB 050712
\citep{depasquale06,lazzati02}, but have been usually linked to a flattening of the X-ray
spectral slope. This is typically an artifact of the spectral fitting routine, that may be 
due to the correlation between spectral parameters such as \nh\ and \bx. 
However,
the situation in GRB\,060729 is completely different. Here we observe not only a
decreasing absorption column density $N_H$, but also a steepening of the X-ray spectral
slope \bx, which is the opposite of what one expects if this is just an artifact of the
fitting routine. Therefore we consider the decrease of the absorption column density to
be real. 
There are two explanations for a decrease in the column density of a neutral absorber: 1)
an expanding medium which results in a lower volume and column density and 2) ionization
of the neutral gas. Both, the expansion and the ionization of the gas, happen after
the explosion of the star. A softening of the X-ray spectrum during the initial
decline has been commonly observed as reported by \citet{zhang_bin07}.

Alternatively the WT mode spectra can also be fitted with a black body plus power law
model. In order to limit the number of free parameters the absorption column density
parameter was set to the Galactic value. These fits show a decrease of the black body
temperature from about 0.6 keV to about 0.1 keV from the beginning  to the end of the WT
observing period. Fixing the absorption column density to a value of $1\times 10^{21}$
cm$^{-2}$, the value obtained at later times during the XMM observation,
results in similar values for the blackbody temperature. From these fits the temperature
tends to be slightly lower than when fixing the $N_H$ to the Galactic value. However,
within the errors the results are consistent. The biggest influence the increase in the
absorption column density has is on the normalization of the blackbody and power law
components. 

In the latter scenario, the thermal component is likely to be the
photospheric emission  from X-ray flares. In the prompt GRB phase, we
have seen the thermal emission from some bursts \citep[e.g. ][]{ryde07},
which has been interpreted as the photospheric emission
when the fireball becomes optically-thin 
\citep{meszaros00, rees05, thompson07, enrico05, peer06}.
Since X-ray flares are believed to be the result of late-time central
engine activity with radiation mechanism similar to the prompt phase, a
photosphere component found in X-ray flares is reasonable.
The thermal emission
component discovered from  GRB060729 also supports the internal
origin of the X-ray flares, rather than external shocks. Let us
examine the relation between the black body radius $R_{bb}$
derived from the spectral fitting and the photosphere radius
$R_{ph}$. For a relativistic moving source, the luminosity of the
thermal component at the photospheric radius is
\begin{equation}
L=4\pi R_{ph}^2\Gamma^2\sigma T'^4
\end{equation}
where $T'$ is the photospheric temperature in the co-moving frame
and $\Gamma$ is the bulk Lorentz factor.  As the usual fitting
uses $L=4\pi R_{bb}^2\sigma T^4$, we get $R_{ph}=\Gamma R_{bb}$ by
taking advantage of $T=\Gamma T'$. The fitted black body radii
around the peak of the X-ray flares are a few times $10^{12}{\rm
cm}$. This means the photospheric radii of the two X-ray flares in
GRB060729 are a few times of $10^{12}\Gamma \,{\rm cm}$. The
photospheric radius is very sensitive to the bulk Lorentz factor
$\Gamma$ of the fireball \citep{rees05} and a
photospheric radius of $10^{13}-10^{14}{\rm cm}$  is reasonable if
$\Gamma$ of the X-ray flare is of the order of ten.

During the early steep decay phase,  the bolometric flux (equivalent
to the flux in XRT band if the peak energy located within the
0.3-10 keV range of the XRT) may decrease as
$F_X(t)\propto(t-t_0)^{-2}$  due to the curvature effect (Ryde \&
Petrosian 2002), where $t_0$ is some reference time of the flare
(Liang et al. 2006), and the peak energy 
decays as $(t-t_0)^{-1}$.  So if we fit the spectrum with the
black body model ($F_X(t)\propto T^4 R_{bb}^2$) all the time
during the decay phase, we would expect that the black body radius
increases with time as $R_{bb}\propto(t-t_0)$. This may explain
the apparent increase of the black body radius with time during
the steep phase of the flares.

\section{Conclusions and Summary}

We studied the \swift\ and \xmm\ observations of the afterglow of GRB\,060729 and found:

\begin{itemize}

\item The light curve of the afterglow extends out to \days days after the
burst (81 days
in the rest-frame) without showing any sign of a jet break. We estimated that the jet
opening angle has to be larger than $28^{\circ}$. This is the longest followup
and detection of a GRB afterglow in X-rays
ever performed, except for GRB 030329.

\item The X-ray light curve can be generally described by an initial steep decay slope
$\alpha_1$=5.1\plm0.2, a break time $T_{\rm break, 1}$=530\plm25 s, flat decay slope
$\alpha_2$=0.14\plm0.02 with a break at $T_{\rm break, 2}$=56.8\plm10 ks, 
and a steep decay slope $\alpha_3$=1.29\plm0.03.

\item The unusually long flat decay phase of the afterglow of GRB\,060729  implies a much
larger energy injection than seen in any other GRB afterglow.

\item After the initial phase, the light curves in X-rays as well as in all 6 UVOT filters
follow the same shape.

\item In the initial phase the afterglow shows a dramatic change in its X-ray spectrum
which can either be described by a steepening of a power law spectrum with a simultaneous
decrease in the intrinsic column density, or a decrease in the black body temperature
from 0.6 keV at 130s after the burst to 0.1 keV at 250s observed after the burst.

\item The spectral analysis of the \swift~XRT PC mode and \xmm\ EPIC pn and MOS data shows
that the X-ray spectrum of the afterglow agrees with an absorbed power law with
\bx=1.1
and an intrinsic column density $N_{\rm H, intr}=1\times 10^{21}$ cm$^{-2}$.

\item The reddening and intrinsic column density estimated from the spectral energy
distribution agrees well with the value found from the \xmm\ analysis.

\end{itemize}

\acknowledgments

We would like to thank Dmitry Frederiks for checking the Konus-Wind measurements of the
time of the burst, and the anonymous referee for a constructive and detailed referee's 
report that significantly improved the paper.
This research has made use of data obtained through the High Energy 
Astrophysics Science Archive Research Center Online Service, provided by the 
NASA/Goddard Space Flight Center.
This research was supported by NASA contract NAS5-00136 and SAO
grant G05-6076X BASIC.

%% Use the figure environment and \plotone or \plottwo to include

%% figures and captions in your electronic submission.

% Table 1

\begin{deluxetable}{lccccccccc}
\rotate
\tabletypesize{\scriptsize}
\tablecaption{Spectral fits to the BAT data. The BAT data have been divided into
5 segments as listed below. \label{bat_spec}
}
\tablewidth{0pt}
\tablehead{
& & & &
\multicolumn{3}{c}{Flux\tablenotemark{1}} &
\multicolumn{3}{c}{Fluence\tablenotemark{2}} \\
\colhead{\rb{spectrum}} &
\colhead{\rb{T after trigger}} &
\colhead{\rb{$\beta_{\rm 15-150 keV}$}} &
\colhead{\rb{$\chi^2/\nu$}} &
\colhead{15--150 keV} &
\colhead{1 keV -- 1 MeV\tablenotemark{3}} &
\colhead{1 keV -- 10 MeV\tablenotemark{3}} &
\colhead{15--150 keV} &
\colhead{1 keV -- 1 MeV\tablenotemark{3}} &
\colhead{1 keV -- 10 MeV\tablenotemark{3}} 
}
\startdata
1st peak  & 0 -- 10 & 1.05$^{+0.42}_{-0.37}$ & 57/56 & 3.2$\times 10^{-8}$     & 1.0$\times 10^{-7}$ & 1.3$\times 10^{-7}$  & 3.2$\times 10^{-7}$ & 1.0$\times 10^{-6}$ & 1.3$\times 10^{-6}$ \\
2nd peak & 70 -- 88 & 0.59$^{+0.11}_{-0.11}$ & 59/56 & 5.8$\times 10^{-8}$    & 1.6$\times 10^{-7}$ & 4.5$\times 10^{-7}$  & 1.0$\times 10^{-6}$ & 2.8$\times 10^{-6}$ & 8.1$\times 10^{-6}$ \\
3rd peak & 88 -- 124 & 0.90$^{+0.11}_{-0.11}$ & 50/56 & 2.8$\times 10^{-8}$   & 7.7$\times 10^{-8}$ & 1.2$\times 10^{-7}$  & 1.0$\times 10^{-6}$ & 2.8$\times 10^{-6}$ & 4.3$\times 10^{-6}$ \\
XRT flare 1 & 124 -- 160 & 1.26$^{+1.58}_{-0.86}$ & 49/56 & 2.6$\times 10^{-9}$& 1.1$\times 10^{-8}$ & 1.3$\times 10^{-8}$  & 9.4$\times 10^{-8}$ & 4.0$\times 10^{-7}$ & 4.7$\times 10^{-7}$ \\
XRT flare 2 & 180 -- 190 & 1.59$^{+5.23}_{-1.67}$ & 53/56 & 1.8$\times 10^{-9}$& 1.5$\times 10^{-8}$ & 1.5$\times 10^{-8}$  & 1.8$\times 10^{-8}$ & 1.5$\times 10^{-7}$ & 1.5$\times 10^{-7}$ 
\enddata

\tablenotetext{1}{15-150 keV flux in units of ergs s$^{-1}$ cm$^{-2}$}
\tablenotetext{2}{Fluence in units of ergs  cm$^{-2}$}
\tablenotetext{3}{Rest-frame 1 keV - 1 MeV (0.65 keV - 650 keV observed) and 1 keV - 10 MeV (0.65 keV - 6.5 MeV observed)}

\end{deluxetable}

% Table 2

%\begin{center}
\begin{deluxetable}{rrrcccccccccccc}
\rotate
\tabletypesize{\tiny}
\tablecaption{\label{wt_xspec} Spectral analysis of  the 21 bins of the 
\swift\ WT mode data.\tablenotemark{1}
}
\tablewidth{0pt}
\tablehead{
 & & &  &
 & \multicolumn{3}{c}{Single Power Law}  
 & \multicolumn{4}{c}{Blackbody + Power Law} 
 & \multicolumn{2}{c}{Power Law with Exponential Cutoff} \\ 
\colhead{\rb{Bin \#}} & 
\colhead{\rb{Time}} & 
\colhead{\rb{$T_{\rm obs}$}} & 
\colhead{\rb{CR\tablenotemark{2}}} & 
\colhead{\rb{HR\tablenotemark{3}}} & 
\colhead{$N_{\rm H}$\tablenotemark{4}} &
\colhead{\bx} & 
\colhead{$\chi^2/\nu$} &
\colhead{kT\tablenotemark{5}} &
\colhead{bb-Flux\tablenotemark{6}} &
\colhead{$R_{\rm bb}$\tablenotemark{7}} &
\colhead{$\chi^2/\nu$} &
\colhead{$E_{\rm cutoff}$\tablenotemark{8}} & 
\colhead{$\chi^2/\nu$} 
} 
\startdata
 1 & 131 & 1.2   & 919\plm30   &  0.55\plm0.06   \\
 2 & 132 & 1.2   & 841\plm28   &  0.60\plm0.06  & \rb{4.06$^{+1.39}_{-1.15}$} & \rb{1.51$^{+0.35}_{-0.31}$} & \rb{21/23} & \rb{0.56$^{+0.05}_{-0.04}$} & 
 \rb{7.46} & \rb{2.49} &  \rb{16/23} &  \rb{0.82$^{+0.44}_{-0.08}$} & \rb{15/23} \\  
 3 & 133 & 1.4   & 807\plm27   &  0.51\plm0.05   \\
 4 & 135 & 1.6   & 737\plm25   &  0.43\plm0.05  & \rb{4.51$^{+0.88}_{-0.77}$} & \rb{2.10$^{+0.29}_{-0.25}$} & \rb{39/32} & \rb{0.45$^{+0.04}_{-0.03}$} & 
 \rb{6.13} & \rb{3.46} &  \rb{51/32} &  \rb{0.78$^{+0.36}_{-0.02}$} & \rb{48/32} \\
 5 & 136 & 1.6   & 712\plm24   &  0.47\plm0.05   \\  
 6 & 138 & 1.8   & 635\plm22   &  0.42\plm0.05  & \rb{5.22$^{+0.98}_{-0.85}$} & \rb{2.40$^{+0.33}_{-0.30}$} & \rb{36/31} & \rb{0.43$^{+0.02}_{-0.02}$} & 
 \rb{6.23} & \rb{3.91} & \rb{34/31} &  \rb{0.51$^{+0.59}_{-0.39}$} & \rb{33/31} \\
 7 & 140 & 2.1   & 579\plm19   &  0.33\plm0.04   \\
 8 & 142 & 2.6   & 520\plm18   &  0.29\plm0.04  & \rb{4.35$^{+0.58}_{-0.53}$} & \rb{2.47$^{+0.24}_{-0.21}$} & \rb{39/46} & \rb{0.38$^{+0.02}_{-0.02}$} & 
 \rb{5.72} & \rb{4.69} & \rb{45/46} &  \rb{0.59$^{+0.46}_{-0.25}$} & \rb{45/46} \\
 9 & 145 & 3.1   & 498\plm16   &  0.25\plm0.04   \\ 
10 & 149 & 4.4   & 380\plm12    &  0.12\plm0.04  & \rb{4.70$^{+0.65}_{-0.58}$} & \rb{2.99$^{+0.32}_{-0.28}$} & \rb{75/50} & \rb{0.31$^{+0.02}_{-0.02}$} &
 \rb{3.60} & \rb{5.48} & \rb{96/50} &  \rb{0.50$^{+0.3}_{-0.2}$} & \rb{107/50} \\
11 & 154 & 5.7   & 289\plm10    &  $-$0.00\plm0.04  & 3.00$^{+0.62}_{-0.56}$ & 2.64$^{+0.34}_{-0.30}$ & 29/34 & 0.25$^{+0.02}_{-0.02}$ 
& 2.81 & 7.35 &  30/34 &  0.67$^{+0.47}_{-0.09}$ & 33/34 \\
12 & 161 & 7.3   & 227\plm7    & $-$0.07\plm0.04  & 3.07$^{+0.71}_{-0.63}$  & 2.73$^{+0.41}_{-0.36}$ & 43/32 & 0.29$^{+0.02}_{-0.02}$ 
& 1.98 & 4.84 & 32/32 &  0.64$^{+0.47}_{-0.15}$ & 34/32 \\ 
13 & 167 & 6.4   & 258\plm9    &  0.10\plm0.04  & 2.29$^{+0.60}_{-0.54}$ & 2.04$^{+0.33}_{-0.29}$ & 38/32 & 0.27$^{+0.03}_{-0.03}$ 
& 1.45 & 4.77 &  43/32 &  1.15$^{+0.64}_{-0.34}$ & 48/32 \\
14 & 173 & 5.2   & 320\plm10    &  0.21\plm0.04  & 3.60$^{+0.62}_{-0.56}$ & 2.46$^{+0.27}_{-0.24}$ & 40/33 & 0.34$^{+0.03}_{-0.02}$ 
& 3.54 & 4.68 & 39/33 &  0.72$^{+0.37}_{-0.05}$ & 37/33 \\
15 & 179 & 5.4   & 310\plm10    &  0.15\plm0.04  & 2.43$^{+0.59}_{-0.53}$ & 2.01$^{+0.30}_{-0.28}$ & 39/35 & 0.32$^{+0.03}_{-0.02}$ 
& 2.90 & 4.61 & 37/35 &  0.78$^{+0.40}_{-0.03}$ & 35/35 \\
16 & 184 & 5.5   & 301\plm10    &  0.02\plm0.04  & 2.97$^{+0.67}_{-0.60}$ & 2.57$^{+0.38}_{-0.34}$ & 35/34 & 0.26$^{+0.02}_{-0.02}$ 
& 3.28 & 7.52 & 32/34 &  0.58$^{+0.56}_{-0.23}$ & 36/34 \\
17 & 190 & 7.4   & 227\plm7    & -0.09\plm0.03  & 3.46$^{+0.85}_{-0.76}$ & 3.07$^{+0.51}_{-0.46}$ & 55/35 & 0.24$^{+0.01}_{-0.01}$ 
& 3.24 & 8.76 & 40/35 &  0.51$^{+0.18}_{-0.13}$ & 45/35 \\ 
18 & 199 & 9.4   & 178\plm5    & -0.26\plm0.03  & 2.48$^{+0.62}_{-0.55}$ & 2.98$^{+0.41}_{-0.36}$ & 62/39 & 0.19$^{+0.01}_{-0.01}$ 
& 3.44 & 14.84 & 58/39 &  0.55$^{+0.62}_{-0.29}$ & 60/39 \\
19 & 210 & 14.0  & 127\plm4 & -0.31\plm0.03  & 2.34$^{+0.63}_{-0.54}$ & 3.07$^{+0.42}_{-0.36}$ & 59/38 & 0.18$^{+0.01}_{-0.01}$ 
& 2.84 & 15.40 & 33/38 &  0.51$^{+0.69}_{-0.31}$ & 60/38 \\
20 & 229 & 23.4  & 72\plm2 & -0.50\plm0.03  & 1.83$^{+0.53}_{-0.44}$ & 3.28$^{+0.41}_{-0.35}$   & 72/36 & 0.15$^{+0.01}_{-0.01}$ 
& 2.15 & 18.19 & 38/36 &  0.51$^{+0.6}_{-0.5}$ & 76/36 \\
21 & 296 & 112.0 & 14.8\plm0.4  & -0.56\plm0.03  & 0.84$^{+0.35}_{-0.30}$ & 2.83$^{+0.36}_{-0.32}$ & 64/37 & 0.11$^{+0.01}_{-0.01}$ 
& 0.50 & 15.65 & 52/37 &  --- & ---
\enddata

\tablenotetext{1}{For the blackbody plus power law and the power law with exponential cutoff model the power law slope was fixed to
\bx=1.0. For the blackbody plus power law model the absorption parameter was fixed to the
Galactic value \citep[4.82$\times 10^{20}$ cm$^{-2}$ ][]{dic90}.}
\tablenotetext{2}{Count rate CR in units of counts s$^{-1}$.}
\tablenotetext{3}{The hardness ratio is defined as $HR=(hard-soft)/(hard+soft)$ 
where $soft$ and $hard$ are the photons in the 0.3-1.0 and
1.0-10.0 keV band, respectively.}
\tablenotetext{4}{The column density $N_{\rm H}$ is given in units of $10^{21}$ cm$^{-2}$.}
\tablenotetext{5}{Blackbody temperature in units of keV.}
\tablenotetext{6}{The fluxes are given in units of 10$^{-9}$ ergs s$^{-1}$ cm$^{-2}$.}
\tablenotetext{7}{Blackbody radius $R_{\rm bb}$ given in units of $10^{12}$ cm.}
\tablenotetext{8}{The break energy is given in units of keV.}

\end{deluxetable}
%\end{center}

% Table 3

\begin{deluxetable}{lcccccc}
\tabletypesize{\scriptsize}
\tablecaption{\label{xmm_xrt_xspec} Spectral fits to the \swift~XRT PC mode and 
XMM EPIC pn and MOS data\tablenotemark{1} 
}
\tablewidth{0pt}
\tablehead{
\colhead{Detector} & 
\colhead{$N_{\rm H}$\tablenotemark{2}} &
\colhead{\bx} & 
\colhead{$\chi^2/\nu$} &
\colhead{$N_{\rm H, intr}${\tablenotemark{3}}} &
\colhead{\bx} & 
\colhead{$\chi^2/\nu$} 
} 
\startdata
\swift~XRT (20--40 ks after burst)   & 16.83$^{+2.30}_{-2.17}$ & 1.21$^{+0.10}_{-0.09}$ & 87/103 & 21.57$^{+4.22}_{-3.92}$ & 1.12\plm0.08 & 88/103 \\
\swift~XRT (44.9--107.5 ks after burst)   & 15.72$^{+2.22}_{-2.10}$ & 1.19$^{+0.09}_{-0.09}$ & 134/120   & 19.20$^{+4.00}_{-3.72}$ & 1.11$^{+0.08}_{-0.07}$ & 134/120 \\
\swift~XRT (200 ks after burst)   & 14.65$^{+6.22}_{-5.55}$ & 1.17$^{+0.25}_{-0.22}$ & 33/25 & 17.46$^{+11.10}_{-9.65}$ & 1.10$^{+0.20}_{-0.19}$ & 33/25 \\
XMM EPIC pn   & 8.58$^{+0.20}_{-0.20}$ & 1.12$^{+0.01}_{-0.01}$ & 1159/1149 &  7.79$^{+0.39}_{-0.38}$ & 1.11$^{+0.01}_{-0.01}$ & 1124/1149 \\
XMM MOS1+MOS2 & 9.33$^{+0.33}_{-0.32}$ & 1.04$^{+0.02}_{-0.02}$ & 937/800   &  8.23$^{+0.59}_{-0.58}$ & 1.01$^{+0.01}_{-0.01}$ & 929/800 \\
XMM MOS 1+2 + \swift~XRT & 9.54$^{+0.32}_{-0.32}$ & 1.05$^{+0.02}_{-0.02}$ & 1106/920 &  8.61$^{+0.06}_{-0.06}$ & 1.02$^{+0.01}_{-0.01}$ & 1097/920 \\
XMM pn + MOS1+2 + \swift~XRT\tablenotemark{3} 
& 8.57$^{+0.17}_{-0.17}$ & 1.08$^{+0.01}_{-0.01}$ & 2599/2071 & 7.55$^{+0.03}_{-0.03}$ & 1.06$^{+0.01}_{-0.01}$ & 2508/2071 
\enddata

\tablenotetext{1}{XMM observed the afterglow between 44.9-107.5 ks after the burst.}
\tablenotetext{2}{The column density $N_{\rm H}$ at z=0 is given in units of $10^{20}$ cm$^{-2}$.}
\tablenotetext{3}{Intrinsic column density $N_{\rm H, intr}$ at the redshift of the burst, z=0.54, is given in units of $10^{20}$ cm$^{-2}$.
The absorption column density at z=0 is set to the Galactic value, 4.82$\times 10^{20}$ cm$^{-2}$.}

\end{deluxetable}

% Table 4

\begin{deluxetable}{rrrcrrcrrcrrcrrcrrc}
\rotate
\tabletypesize{\tiny}
\tablecaption{\label{uvot_lc} Central times, exposure times and aperture corrected 
magnitudes of the UVOT light curves
}
\tablewidth{0pt}
\tablehead{
 & \multicolumn{3}{c}{V-Filter}  
 & \multicolumn{3}{c}{B-Filter} 
 & \multicolumn{3}{c}{U-Filter} 
 & \multicolumn{3}{c}{UVW1-Filter} 
 & \multicolumn{3}{c}{UVM2-Filter} 
 & \multicolumn{3}{c}{UVW2-Filter} \\
\colhead{\rb{Bin \#}} & 
\colhead{Time\tablenotemark{1}} & 
\colhead{$T_{\rm exp}$\tablenotemark{2}} & 
\colhead{Mag} & 
\colhead{Time\tablenotemark{1}} & 
\colhead{$T_{\rm exp}$\tablenotemark{2}} & 
\colhead{Mag} & 
\colhead{Time\tablenotemark{1}} & 
\colhead{$T_{\rm exp}$\tablenotemark{2}} & 
\colhead{Mag} & 
\colhead{Time\tablenotemark{1}} & 
\colhead{$T_{\rm exp}$\tablenotemark{2}} & 
\colhead{Mag} & 
\colhead{Time\tablenotemark{1}} & 
\colhead{$T_{\rm exp}$\tablenotemark{2}} & 
\colhead{Mag} & 
\colhead{Time\tablenotemark{1}} & 
\colhead{$T_{\rm exp}$\tablenotemark{2}} & 
\colhead{Mag} 
}
\startdata
 1 & 120    & 9    & 17.07\plm0.31 & 715    & 10   & 18.27\plm0.38 & 696    & 20   & 16.48\plm0.12 & 673     & 20    & 16.92\plm0.20 & 649    & 18   & 16.81\plm0.24 & 749    & 20     & 17.54\plm0.25 \\
 2 & 440    & 390  & 17.37\plm0.06 & 1451   & 20   & 18.03\plm0.23 & 844    & 20   & 16.88\plm0.15 & 820     & 20    & 16.89\plm0.20 & 796    & 20   & 17.57\plm0.33 & 1489   & 20     & 16.95\plm0.19 \\
 3 & 773    & 19   & 17.36\plm0.26 & 1609   & 20   & 18.10\plm0.26 & 1427   & 20   & 16.85\plm0.16 & 1403    & 20    & 17.47\plm0.26 & 1379   & 20   & 17.12\plm0.26 & 1647   & 20     & 17.39\plm0.23 \\
 4 & 1164   & 393  & 17.32\plm0.06 & 1767   & 20   & 18.26\plm0.30 & 1585   & 20   & 16.85\plm0.15 & 1561    & 20    & 17.31\plm0.24 & 1537   & 20   & 16.85\plm0.23 & 1805   & 19     & 16.92\plm0.19 \\
 5 & 1513   & 19   & 17.45\plm0.32 & 6224   & 197  & 18.45\plm0.10 & 1743   & 20   & 16.49\plm0.13 & 1719    & 20    & 17.14\plm0.23 & 1695   & 20   & 16.75\plm0.22 & 6634   & 197    & 17.70\plm0.09 \\
 6 & 1671   & 19   & 17.16\plm0.24 & 7657   & 197  & 18.45\plm0.11 & 6020   & 197  & 17.52\plm0.07 & 5815    & 197   & 17.55\plm0.09 & 1850   & 13   & 17.63\plm0.45 & 7998   & 63     & 17.86\plm0.17 \\
 7 & 1829   & 19   & 17.55\plm0.34 & 12919  & 134  & 18.68\plm0.14 & 7452   & 197  & 17.54\plm0.07 & 7247    & 197   & 17.62\plm0.09 & 7043   & 197  & 17.80\plm0.12 & 11878  & 751    & 18.08\plm0.06 \\
 8 & 6838   & 197  & 18.10\plm0.16 & 18042  & 211  & 17.86\plm0.06 & 12778  & 134  & 17.89\plm0.11 & 12568   & 268   & 18.09\plm0.10 & 13880   & 377  & 17.76\plm0.09 & 13266  & 537	& 18.17\plm0.07 \\
 9 & 13613  & 134  & 18.38\plm0.25 & 23824  & 211  & 17.84\plm0.06 & 17822  & 211  & 16.91\plm0.05 & 17495   & 422   & 17.11\plm0.05 & 19550  & 600  & 17.07\plm0.05 & 18585  & 844    & 17.36\plm0.04 \\
10 & 19128  & 211  & 17.33\plm0.09 & 29606  & 211  & 17.78\plm0.06 & 23603  & 211  & 17.20\plm0.06 & 23276   & 422   & 17.23\plm0.05 & 25338  & 599  & 17.20\plm0.05 & 24367  & 845    & 17.52\plm0.04 \\
11 & 24911  & 211  & 17.73\plm0.11 & 35388  & 211  & 18.14\plm0.08 & 29387  & 211  & 17.17\plm0.06 & 29059   & 422   & 17.22\plm0.05 & 31109  & 600  & 17.37\plm0.06 & 30149  & 844    & 17.51\plm0.04 \\
12 & 30692  & 211  & 17.53\plm0.10 & 41205  & 208  & 18.12\plm0.08 & 35167  & 211  & 17.20\plm0.06 & 34840   & 422   & 17.35\plm0.05 & 36892  & 599  & 17.40\plm0.06 & 35931  & 845    & 17.64\plm0.04 \\
13 & 36475  & 211  & 17.98\plm0.13 & 47494  & 147  & 18.22\plm0.09 & 40989  & 207  & 17.26\plm0.06 & 40666   & 415   & 17.35\plm0.05 & 42681  & 587  & 17.45\plm0.06 & 41738  & 829    & 17.76\plm0.05 \\
14 & 42273  & 207  & 17.86\plm0.13 & 56873  & 129  & 18.31\plm0.11 & 47340  & 147  & 17.18\plm0.07 & 47109   & 296   & 17.24\plm0.06 & 48550  & 416  & 17.25\plm0.06 & 47877  & 592    & 17.56\plm0.05 \\
15 & 48259  & 147  & 17.47\plm0.11 & 67943  & 42   & 18.94\plm0.33 & 53746  & 83   & 17.32\plm0.10 & 52836   & 183   & 17.36\plm0.08 & 54427  & 229  & 17.63\plm0.10 & 54050  & 330    & 17.78\plm0.08 \\
16 & 54265  & 83   & 18.12\plm0.26 & 75864  & 211  & 18.77\plm0.12 & 62872  & 69   & 17.24\plm0.11 & 58802   & 385   & 17.60\plm0.06 & 63155  & 177  & 17.93\plm0.14 & 62989  & 277    & 18.18\plm0.10 \\
17 & 63099  & 69   & 18.29\plm0.35 & 81646  & 211  & 18.88\plm0.13 & 69860  & 211  & 17.75\plm0.08 & 64698   & 436   & 17.76\plm0.07 & 100500 & 603  & 18.66\plm0.11 & 76105  & 249    & 18.14\plm0.10 \\
18 & 91349  & 223  & 18.68\plm0.22 & 87426  & 211  & 18.96\plm0.14 & 75644  & 211  & 17.84\plm0.08 & 69533   & 422   & 17.85\plm0.07 & 106280 & 599  & 18.57\plm0.11 & 82086  & 640    & 18.48\plm0.08 \\
19 & 100080 & 211  & 18.45\plm0.22 & 93208  & 211  & 18.72\plm0.11 & 81426  & 211  & 18.07\plm0.10 & 75315   & 423   & 17.93\plm0.07 & 112060 & 597  & 18.58\plm0.11 & 87970  & 846    & 18.54\plm0.07 \\
20 & 105860 & 211  & 18.87\plm0.28 & 98988  & 211  & 19.11\plm0.15 & 87206  & 211  & 17.98\plm0.10 & 81097   & 423   & 18.05\plm0.08 & 117840 & 600  & 18.52\plm0.10 & 93752  & 846    & 18.84\plm0.08 \\
21 & 111650 & 211  & 18.49\plm0.19 & 104770 & 211  & 19.02\plm0.13 & 92987  & 211  & 18.03\plm0.10 & 86877   & 423   & 18.17\plm0.08 & 123630 & 600  & 18.90\plm0.13 & 99532  & 847    & 18.79\plm0.08 \\
22 & 117430 & 211  & 19.03\plm0.32 & 110560 & 211  & 18.84\plm0.13 & 98767  & 211  & 18.40\plm0.12 & 92658   & 423   & 18.35\plm0.09 & 129410 & 601  & 18.81\plm0.12 & 105320 & 846    & 18.68\plm0.07 \\
23 & 123210 & 211  & 18.40\plm0.20 & 116340 & 211  & 19.18\plm0.16 & 104550 & 211  & 18.23\plm0.11 & 98438   & 423   & 18.33\plm0.09 & 135340 & 294  & 18.79\plm0.17 & 111100 & 846    & 18.74\plm0.08 \\
24 & 128990 & 211  & 18.92\plm0.27 & 122120 & 211  & 19.62\plm0.23 & 110340 & 211  & 18.50\plm0.14 & 104230  & 423   & 18.53\plm0.10 & 146960 & 342  & 18.99\plm0.19 & 116880 & 846    & 18.87\plm0.08 \\
25 & 135120 & 111  & 18.77\plm0.38 & 127900 & 211  & 19.32\plm0.17 & 116120 & 211  & 18.33\plm0.12 & 110010  & 423   & 18.52\plm0.10 & 195280 & 737  & 19.62\plm0.20 & 122660 & 845    & 19.05\plm0.09 \\
26 & 158090 & 156  & 19.36\plm0.56 & 140600 & 225  & 19.37\plm0.19 & 121900 & 211  & 18.60\plm0.16 & 115790  & 423   & 18.56\plm0.10 & 204280 & 189  & 19.48\plm0.11 & 128450 & 845    & 19.05\plm0.09 \\
27 & 226950 & 941  & 19.82\plm0.29 & 154780 & 180  & 19.77\plm0.30 & 127680 & 211  & 18.43\plm0.12 & 121570  & 422   & 18.63\plm0.11 & 235840 & 471  & 19.84\plm0.29 & 134830 & 445    & 19.07\plm0.13 \\
28 & 319820 & 1374 & 20.22\plm0.34 & 168380 & 211  & 19.66\plm0.24 & 134430 & 110  & 18.46\plm0.17 & 127360  & 423   & 18.76\plm0.11 & 288290 & 1550 & 19.86\plm0.16 & 146750 & 522    & 18.84\plm0.11 \\
29 & 576790 & 5531 & 21.04\plm0.40 & 174170 & 209  & 20.00\plm0.32 & 146620 & 178  & 19.08\plm0.25 & 137350  & 412   & 18.79\plm0.12 & 331760 & 2240 & 20.21\plm0.18 & 168860 & 729    & 19.32\plm0.11 \\
30 & 897975 & 2368 & 3$\sigma$ul=20.92 & 179940 & 211  & 20.00\plm0.32 & 156590 & 211  & 18.64\plm0.15 & 145550  & 291   & 18.76\plm0.14 & 397970 & 3140 & 20.33\plm0.16 & 174710 & 839& 19.20\plm0.10 \\
31 &\nodata &\nodata & \nodata     & 194290 & 275  & 19.68\plm0.20 & 168160 & 211  & 18.70\plm0.15 & 150480  & 423   & 18.78\plm0.12 & 487880 & 2980 & 20.59\plm0.20 & 180340 & 556    & 19.31\plm0.13 \\
32 &\nodata &\nodata & \nodata     & 206460 & 273  & 19.29\plm0.15 & 173950 & 209  & 18.84\plm0.16 & 154560  & 456   & 18.94\plm0.12 & 663250 & 8830 & 21.68\plm0.30 & 194650 & 1100   & 19.60\plm0.11 \\
33 &\nodata &\nodata & \nodata     & 319120 & 1374 & 20.11\plm0.14 & 179720 & 211  & 19.09\plm0.21 & 162860  & 261   & 19.18\plm0.19 & 926689 & 3976 & 3$\sigma$ul=21.96 & 206810 &1092& 19.51\plm0.10 \\
34 &\nodata &\nodata & \nodata     & 449360 & 2304 & 20.38\plm0.13 & 194150 & 275  & 18.82\plm0.13 & 167830  & 423   & 19.04\plm0.13 &\nodata &\nodata & \nodata  & 218500 & 8795   & 19.52\plm0.12 \\
35 &\nodata &\nodata & \nodata     & 662500 & 3619 & 21.23\plm0.20 & 206310 & 273  & 19.03\plm0.16 & 173630  & 420   & 18.98\plm0.13 &\nodata &\nodata & \nodata  & 250410 & 1111   & 19.84\plm0.13 \\
36 &\nodata &\nodata & \nodata     & 897470 & 2368 & 21.48\plm0.29 & 221160 & 280  & 19.35\plm0.22 & 179390  & 423   & 19.20\plm0.15 &\nodata &\nodata & \nodata  & 287670 & 2200   & 20.16\plm0.11 \\
37 &\nodata &\nodata & \nodata	   &\nodata &	\nodata &\nodata    & 253230 & 361  & 19.57\plm0.23 & 193930  & 549   & 19.21\plm0.12 &\nodata &\nodata & \nodata  & 331140 & 3288   & 20.59\plm0.13 \\
38 &\nodata &\nodata & \nodata	    &\nodata &	\nodata &\nodata    & 318970 & 1374 & 19.75\plm0.13 & 206100  & 545   & 19.37\plm0.14 &\nodata &\nodata & \nodata  & 397340 & 4561   & 20.57\plm0.11 \\
39 &\nodata &\nodata & \nodata	    &\nodata &	\nodata &\nodata    & 449220 & 2304 & 20.39\plm0.16 & 220980  & 559   & 19.37\plm0.15 &\nodata &\nodata & \nodata  & 487280 & 4656   & 20.99\plm0.14 \\
40 &\nodata &\nodata & \nodata	    &\nodata &	\nodata &\nodata    & 662390 & 3622 & 21.55\plm0.35 & 286960  & 1100  & 19.77\plm0.14 &\nodata &\nodata & \nodata  & 662770 & 13555  & 21.45\plm0.12 \\
41 &\nodata &\nodata & \nodata	    &\nodata &	\nodata &\nodata    & 897363 & 2368 &3$\sigma$ul=21.58 & 330430  & 1648  & 19.93\plm0.13& \nodata & \nodata & \nodata & 897780 & 9498	& 22.02\plm0.23 \\
42 &\nodata &\nodata & \nodata	    &\nodata &  \nodata &\nodata    &\nodata &  \nodata &\nodata & 396620  & 2276  & 20.09\plm0.11 &\nodata &  \nodata &\nodata   & 1099600& 13675  & 22.43\plm0.29 \\
43 &\nodata &\nodata & \nodata	    &\nodata &  \nodata &\nodata    &\nodata &  \nodata &\nodata & 486600  & 2323  & 20.29\plm0.13 &\nodata &  \nodata &\nodata   & 1487290 & 17912 & 3$\sigma$ul=23.04 \\
44 &\nodata &\nodata & \nodata	    &\nodata &  \nodata &\nodata    &\nodata &  \nodata &\nodata & 662230  & 7331  & 21.20\plm0.15  &\nodata &  \nodata &\nodata  &\nodata &  \nodata &\nodata   \\
45 &\nodata &\nodata & \nodata	    &\nodata &  \nodata &\nodata    &\nodata &  \nodata &\nodata & 897220  & 4748  & 21.20\plm0.18  &\nodata &  \nodata &\nodata  &\nodata &  \nodata &\nodata    \\
46 &\nodata &\nodata & \nodata	    &\nodata &  \nodata &\nodata    &\nodata &  \nodata &\nodata & 1012900 & 17014 & 21.65\plm0.15  &\nodata &  \nodata &\nodata  &\nodata &  \nodata &\nodata    \\
47 &\nodata &\nodata & \nodata	    &\nodata &  \nodata &\nodata    &\nodata &  \nodata &\nodata & 1614900 & 12901 & 21.82\plm0.22  &\nodata &  \nodata &\nodata  &\nodata &  \nodata &\nodata    \\
48 &\nodata &\nodata & \nodata	    &\nodata &  \nodata &\nodata    &\nodata &  \nodata &\nodata & 1697900 & 14706 & 22.46\plm0.32  &\nodata &  \nodata &\nodata  &\nodata &  \nodata &\nodata    \\
49 &\nodata &\nodata & \nodata	    &\nodata &  \nodata &\nodata    &\nodata &  \nodata &\nodata & 1874500 & 24764 & 22.68\plm0.27  &\nodata &  \nodata &\nodata  &\nodata &  \nodata &\nodata    \\
50 &\nodata &\nodata & \nodata	    &\nodata &  \nodata &\nodata    &\nodata &  \nodata &\nodata & 2091400 & 20552 & 22.11\plm0.20  &\nodata &  \nodata &\nodata  &\nodata &  \nodata &\nodata    \\
51 &\nodata &\nodata & \nodata	    &\nodata &  \nodata &\nodata    &\nodata &  \nodata &\nodata & 2394800 & 39794 & 22.63\plm0.38   &\nodata &  \nodata &\nodata  &\nodata &  \nodata &\nodata   \\
52 &\nodata &\nodata & \nodata      &\nodata & \nodata  &\nodata    &\nodata &\nodata & \nodata	   & 3431882 & 44673 &$\sigma$ul=22.76 &\nodata & \nodata & \nodata &\nodata &  \nodata & \nodata
\enddata

\tablenotetext{1}{The times mark the middle of the time bin in s after the burst.}
\tablenotetext{2}{The exposure times $T_{\rm exp}$ are given in s.}	 
	   								 
\end{deluxetable}

%table 5
	   								 
\begin{deluxetable}{cccc}						 
\tabletypesize{\scriptsize}						 
\tablecaption{\label{uvot_lc_slope} Decay slopes\tablenotemark{1} 
and break times of the UVOT light curves. 
}	   
\tablewidth{0pt}
\tablehead{
\colhead{Filter} &
\colhead{$\alpha_{\rm 2}$\tablenotemark{2}} &
\colhead{$T_{\rm break}$\tablenotemark{3}} &
\colhead{$\alpha_{\rm 3}$}
} 	   
\startdata 
V     & 0.24\plm0.05 & 40$^{+15}_{-15}$ & 1.21\plm0.09 \\
B     & 0.16\plm0.05 & 30$^{+20}_{-10}$ & 0.96\plm0.05\tablenotemark{4} \\
U     & 0.41\plm0.07 & 54$^{+10}_{-15}$ & 1.40\plm0.07 \\
UVW1 & 0.47\plm0.06 & 48$^{+5}_{-5}$	& 1.29\plm0.03 \\
UVM2 & 0.35\plm0.10 & 50$^{+5}_{-15}$  & 1.39\plm0.05 \\
UVW2 & 0.46\plm0.07 & 55$^{+5}_{-10}$  & 1.36\plm0.04 \\
X-rays & 0.14\plm0.02 & 57$^{+10}_{-10}$ & 1.29\plm0.03
\enddata   

\tablenotetext{1}{The decay slopes $\alpha_{\rm 2}$ and $\alpha_{\rm 3}$ follow
the conventions in \citet{zhang06} and \citet{nousek06}.}
\tablenotetext{2}{Decay slope before the energy injection and rebrightening at
20 ks after the burst}
\tablenotetext{3}{Time after the break in ks in the observed frame}
\tablenotetext{4}{Note that when limiting the analysis of the late time decay slope
$\alpha_3$ to 50-200 ks after the burst, the decay slope is 1.17\plm0.16.}

\end{deluxetable}

% Table 6
	   
\begin{deluxetable}{rcc}
\tabletypesize{\scriptsize}
\tablecaption{\label{uvot_early} List of the \swift~UVOT White and V 
observations during the first orbit as shown in
Figure\,\ref{grb060729_xrt_uvot_early}.
}	   
\tablewidth{0pt}
\tablehead{
\colhead{Time\tablenotemark{1}} &
\colhead{White} &
\colhead{V} 
} 	   
\startdata 
  140 &  17.27\plm0.18  & \nodata \\
  150 &  17.51\plm0.21  & \nodata \\
  160 &  17.93\plm0.27  & \nodata \\ 
  170 &  17.86\plm0.26  & \nodata \\ 
  180 &  18.01\plm0.30  & \nodata \\ 
  190 &  3$\sigma$ul: 18.37 & \nodata \\
  200 &  3$\sigma$ul: 18.36 & \nodata \\
  210 &  17.69\plm0.23  & \nodata \\ 
  220 &  18.17\plm0.35  & \nodata \\ 
  230 &  17.44\plm0.21  & \nodata \\ 
  278 & \nodata & 17.50\plm0.30 \\ 
  303 & \nodata & 3$\sigma$ul: 17.81 \\
  328 & \nodata & 17.32\plm0.26 \\ 
  353 & \nodata & 3$\sigma$ul: 17.70 \\
  378 & \nodata & 3$\sigma$ul: 17.79 \\
  403 & \nodata & 17.17\plm0.24 \\ 
  428 & \nodata & 16.99\plm0.22 \\ 
  453 & \nodata & 17.03\plm0.22 \\ 
  478 & \nodata & 17.26\plm0.27 \\ 
  503 & \nodata & 17.12\plm0.23 \\ 
  528 & \nodata & 16.88\plm0.19 \\ 
  553 & \nodata & 17.42\plm0.31 \\ 
  578 & \nodata & 17.20\plm0.26 \\ 
  603 & \nodata & 17.46\plm0.30 \\ 
  628 & \nodata & 16.89\plm0.21 \\ 
  778 & \nodata & 17.27\plm0.28 \\ 
  978 & \nodata & 17.59\plm0.34 \\ 
 1003 & \nodata & 17.66\plm0.33 \\ 
 1028 & \nodata & 17.55\plm0.35 \\ 
 1053 & \nodata & 16.94\plm0.19 \\ 
 1078 & \nodata & 17.60\plm0.32 \\ 
 1103 & \nodata & 17.02\plm0.21 \\ 
 1128 & \nodata & 17.40\plm0.29 \\ 
 1153 & \nodata & 3$\sigma$ul: 17.73 \\
 1178 & \nodata & 17.37\plm0.27 \\ 
 1203 & \nodata & 16.97\plm0.20 \\ 
 1228 & \nodata & 17.43\plm0.29 \\ 
 1253 & \nodata & 16.86\plm0.19 \\ 
 1278 & \nodata & 3$\sigma$ul: 17.70 \\
 1303 & \nodata & 16.94\plm0.19 \\ 
 1328 & \nodata & 17.42\plm0.29 \\ 
 1353 & \nodata & 17.08\plm0.22 \\ 
 1503 & \nodata & 3$\sigma$ul: 17.19 \\
 1528 & \nodata & 3$\sigma$ul: 17.81 \\
 1678 & \nodata & 16.96\plm0.27 \\ 
\enddata   

\tablenotetext{1}{The times note the middle of the observation in s after the burst}

\end{deluxetable}

% Table 7
	   
\begin{deluxetable}{crcc}
\tabletypesize{\scriptsize}
\tablecaption{\label{om_data} List of the \xmm~Optical Monitor observations
}	   
\tablewidth{0pt}
\tablehead{
\colhead{ObsID} &
\colhead{$T_{\rm obs}$\tablenotemark{1}} &
\colhead{Filter} &
\colhead{Mag}
} 	   
\startdata 
006 &  48351 & U    & 17.27\plm0.02 \\
018 &  52707 & U    & 17.43\plm0.02 \\
010 &  70984 & UVW1 & 17.53\plm0.04 \\
015 &  75291 & UVW1 & 17.57\plm0.04 \\
016 &  84105 & UVW1 & 17.71\plm0.04 \\
020 &  88612 & UVW1 & 17.83\plm0.05 \\
012 &  92919 & UVM2 & 17.89\plm0.12 \\
014 &  97226 & UVM2 & 17.98\plm0.13 \\
017 & 101533 & UVM2 & 17.73\plm0.11   
\enddata   

\tablenotetext{1}{The times note the middle of the observation in s after the burst}

\end{deluxetable}

\end{document}